\documentclass{pasj02}

\usepackage[whole]{bxcjkjatype}
\usepackage{natbib}
\usepackage{url}
\usepackage[switch,mathlines]{lineno}

\def\arcdeg{$^\circ$}

\Received{yyyy/mm/dd}
\Accepted{yyyy/mm/dd}
\Published{yyyy/mm/dd}

\newcommand{\nar}{New Astron. Rev.}
\begin{document} 

\title{
Bar-Informed Kinematic-Distance Mapping of Molecular Gas in the Inner Milky Way
}

\author{Junichi \textsc{Baba}\altaffilmark{1,2}\altemailmark \orcid{0000-0002-2154-8740}}
\email{babajn2000@gmail.com; junichi.baba@sci.kagoshima-u.ac.jp}

\altaffiltext{1}{Amanogawa Galaxy Astronomy Research Center, Graduate School of Science and Engineering, Kagoshima University, 1-21-35 Korimoto, Kagoshima 890-0065, Japan.}
\altaffiltext{2}{Division of Science, National Astronomical Observatory of Japan, Mitaka, Tokyo 181-8588, Japan.}

\KeyWords{Galaxy: structure --- Galaxy: bulge --- ISM: kinematics and dynamics --- methods: numerical --- radio lines: ISM}

\maketitle

\begin{abstract}
We present a bar-informed kinematic-distance (BIKD) method to reconstruct face-on molecular-gas maps of the inner Milky Way from position--position--velocity data, relaxing the standard assumption of axisymmetric circular rotation that can generate severe artifacts in barred regions. 
BIKD replaces the rotation curve with a non-axisymmetric streaming field extracted from hydrodynamical simulations in an observationally constrained barred Galactic potential, and infers a discrete distance posterior along each sightline using a Gaussian likelihood in line-of-sight velocity. 
To mitigate multi-modality, we adopt posterior-weighted map making via posterior sampling (a soft-assignment scheme), including a reweighting to reduce velocity-crowding bias. 
We validate the full pipeline in closed-loop tests on the simulations, showing that the recovered large-scale morphology is only weakly sensitive to simple distance priors and remains stable across plausible variations in bar angle, snapshot time, and pattern speed. 
We then apply BIKD to a Galactic CO(1--0) survey to obtain a face-on $\Sigma_{\mathrm{H_2}}$ map. 
Compared to a standard axisymmetric kinematic-distance (KD) reconstruction, BIKD strongly suppresses line-of-sight--elongated finger-of-God features and robustly recovers a bar-aligned, quadrant-asymmetric inner-Galaxy morphology under model marginalization. 
The model-marginalized radial profiles show an approximately exponential decline beyond $\sim4$~kpc, a pronounced deficit at $R\sim0.5$--$3.5$~kpc (a bar gap), and a central concentration consistent with Central Molecular Zone surface densities. 
Finally, we compare prominent ridge-shaped overdensities in the BIKD map with independent spiral-arm loci traced by high-mass star-forming region masers with very long baseline interferometry (VLBI) trigonometric parallaxes and by classical Cepheids with period--luminosity distances. 
Several maser-parallax segments are qualitatively consistent with the dominant BIKD ridges, whereas the Cepheid loci do not coincide with them within their recommended azimuth range. 
Overall, BIKD provides a practical, model-marginalized face-on reconstruction of molecular gas in the inner Milky Way that can be compared directly with modern stellar maps.
\end{abstract}


\section{Introduction}
\label{sec:intro}

The Milky Way is now entering an era in which the stellar component is mapped not only in three-dimensional space but also in its dynamical phase-space structure with unprecedented fidelity \citep[for a review,][]{HuntVasiliev2025}.
Near-infrared surface-brightness maps and red-clump star counts have long provided direct evidence for a triaxial bulge/bar and enabled three-dimensional density reconstructions of the inner Milky Way \citep[e.g.,][]{Dwek+1995,Binney+1997,WeggGerhard2013,Wegg+2015}.
In the \textit{Gaia} era, space-based astrometry combined with large ground-based photometric and spectroscopic surveys has further enabled detailed, global reconstructions of the stellar density and three-dimensional extinction, revealing the Galactic bar directly in map form \citep[e.g.,][]{Anders+2019,Marshall+2025}.
In contrast, the spatial distribution of interstellar gas remains substantially more uncertain because direct distance measurements are generally impractical for extended emission, unlike for point-like sources \citep[e.g.][]{GaiaCollaboration2016,ReidHonma2014}.
Bridging this gap between the well-mapped stellar bar and the poorly constrained gas distribution is essential for a coherent dynamical picture of the inner Milky Way.

Observationally, Galactic gas is measured in position--position--velocity (PPV) space, i.e., Galactic longitude, latitude, and line-of-sight velocity $(l,b,v)$, for example via H\,\textsc{i} and CO line surveys \citep[e.g.,][]{Dame+2001,Kalberla+2005,Jackson+2006,HI4PICollab+2016,HeyerDame2015}. Such surveys provide a three-dimensional data cube $I(l,b,v)$ of observed line intensity. Gas dynamical models in a barred Galactic potential predict the gas surface density and streaming motions in face-on $(X,Y)$ space, so confronting them with the data requires mapping $I(l,b,v)$ into three-dimensional positions along each line of sight. Because $v$ is not a direct distance indicator, this mapping is not unique without an additional kinematic model.

The standard ``kinematic-distance'' (KD) method deprojects PPV data by assuming axisymmetric circular rotation and inverting the line-of-sight velocity--distance relation along each sightline \citep[][]{Oort+1958}, and it has been widely used to produce face-on H\,\textsc{i}/CO maps \citep[][]{NakanishiSofue2003,NakanishiSofue2006,NakanishiSofue2016,Roman-Duval+2009,Marasco+2017,Sofue2023}.
However, this axisymmetric assumption is strongly violated in the barred inner Milky Way: the gas kinematics are inconsistent with circular motion in an axisymmetric potential and exhibit strong bar-driven non-circular streaming \citep[e.g.,][]{Binney+1991,EnglmaierGerhard1999,WeinerSellwood1999,Fux1999,Bissantz+2003,Rodriguez-FernandezCombes2008,Baba+2010,Sormani+2015a}.
Applying an axisymmetric KD inversion to such non-circular flows makes the velocity--distance mapping non-monotonic, so the inversion becomes non-unique (multiple distances can correspond to the same line-of-sight velocity), leading to severe ambiguities and large-scale artifacts in reconstructed maps \citep[][]{Pohl+2008,Hunter+2024,Baba+2009,Baba2026a}.
As a result, it is unclear to what extent prominent structures in KD-based maps reflect true gas density features versus kinematic projection effects.
More generally, even with a kinematic model, converting PPV data into three-dimensional positions is often ill-posed and can yield no solution or multiple distance solutions along a given sightline.

In recent years, the gravitational potential of the inner Milky Way has been constrained by stellar dynamical modeling that incorporates bulge/bar kinematics and photometry, for example via made-to-measure (M2M) approaches \citep[][]{Portail+2017}, which adjust the weights of an $N$-body particle model so that it reproduces the observed density and kinematic constraints while remaining dynamically consistent \citep[][]{SeyerTremaine1996}.
These models provide observationally motivated bar geometries and pattern speeds and thus enable physically grounded predictions of gas streaming motions in a barred potential when coupled to hydrodynamical simulations \citep[][]{Li+2016,Li+2022,Hunter+2024,Baba2025b}.
This motivates an alternative approach to distance inference: rather than enforcing axisymmetric circular rotation, one can use a non-axisymmetric, bar-informed streaming model to interpret observed PPV data.

Bar-informed deprojection is not a new idea. \citet{Pohl+2008} showed that a hydrodynamic model in a barred potential can replace a circular rotation curve with a non-axisymmetric distance--velocity relation, improving inner-Galaxy CO deprojections, while noting that residual line-of-sight stretching can persist if the adopted flow model is imperfect or if distance ambiguities are not handled carefully. More recently, \citet{MertschVittino2021} and \citet{MertschPhan2023} cast the PPV-to-3D problem as Bayesian field inference, reconstructing three-dimensional gas-density fields under spatial-coherence priors that regularize the ill-posed inversion while adopting prescribed gas-flow models that include bar-driven non-circular motions. In this formulation, distance ambiguities are absorbed into the inference and manifest as posterior uncertainties in the recovered density field. A remaining limitation is that the velocity field is still assumed; \citet{Soding+2025} address this by extending the framework to jointly infer the gas density and a spatially varying velocity field.
Taken together, these studies highlight two recurring challenges in the barred inner Milky Way: distance solutions are intrinsically multi-valued along many sightlines, often yielding multi-modal distance posteriors, and the inferred morphology can depend sensitively on the adopted streaming field.

In this paper we develop and test a ``bar-informed kinematic-distance'' (BIKD) method tailored to (i) streaming fields derived from hydrodynamical simulations in observationally constrained barred Milky Way potentials and 
(ii) map making that remains stable in the presence of strongly multi-modal distance posteriors by propagating line-of-sight distance ambiguities into the final maps.
BIKD replaces the axisymmetric rotation model with a spatially varying non-axisymmetric velocity field and infers a distance posterior $P(s\mid l,b,v,\mathcal{M})$, where $s$ is the heliocentric distance and $\mathcal{M}$ labels the adopted streaming model.
Because $P(s\mid l,b,v,\mathcal{M})$ can be strongly multi-modal in the inner Milky Way, we construct face-on maps by posterior-weighted map making (posterior sampling; a soft-assignment scheme), explicitly propagating line-of-sight distance ambiguities into the final maps.
Finally, by repeating the reconstruction over an ensemble of plausible barred streaming models, we quantify the sensitivity of the inferred morphology to bar parameters and time variability and provide model-marginalized maps and uncertainty estimates.

The paper is organized as follows.
Section~\ref{sec:BIKDmethod} introduces the BIKD formalism, including the bar-informed streaming model, the line-of-sight distance posterior, and posterior-sampling (soft-assignment) map making with a velocity-crowding reweighting.
Section~\ref{sec:validation} presents closed-loop validation tests on simulation data and assesses systematic effects associated with bar angle, snapshot-to-snapshot variability, and pattern speed, motivating our model-marginalized treatment.
Section~\ref{sec:obs} applies BIKD to the Dame et al.\ CO(1--0) survey and presents the resulting face-on $\Sigma_{\mathrm{H_2}}$ map, including comparisons to a standard axisymmetric KD reconstruction, model-marginalized profiles, and comparisons with literature spiral-arm loci.
Finally, we summarize our conclusions in Section~\ref{sec:summary}.

\section{Bar-informed kinematic distance method}
\label{sec:BIKDmethod}

BIKD reconstructs a face-on gas map from PPV information by combining (i) a bar-informed streaming model and (ii) probabilistic distance inference along each line of sight.
The inputs are (a) an observational PPV cube (or, for validation, a set of simulation particles with known $(l,b,v)$) and (b) a non-axisymmetric velocity field $\bar{\mathbf{v}}$ and associated dispersion field $\sigma_{\rm los}$ derived from hydrodynamical simulations in a barred Milky Way potential \citep[][]{Baba2025b}. 

Throughout this paper, $(l,b)$ denote Galactic longitude and latitude in the Galactic coordinate system, and $v\equiv v_{\rm los}$ is the line-of-sight velocity with respect to the local standard of rest (LSR), taken to be positive for recession.
We adopt a Sun--Galactic-center distance $R_0=8.3~\mathrm{kpc}$ and an LSR circular speed $V_0=238~\mathrm{km\,s^{-1}}$ at the Solar radius \citep[][]{Bland-HawthornGerhard2016,GravityCollaboratio+2021}.

\subsection{Bar-informed streaming model}
\label{subsec:BIKD_geometry_vfield}

\begin{figure*}
\begin{center}
\includegraphics[width=1.\textwidth]{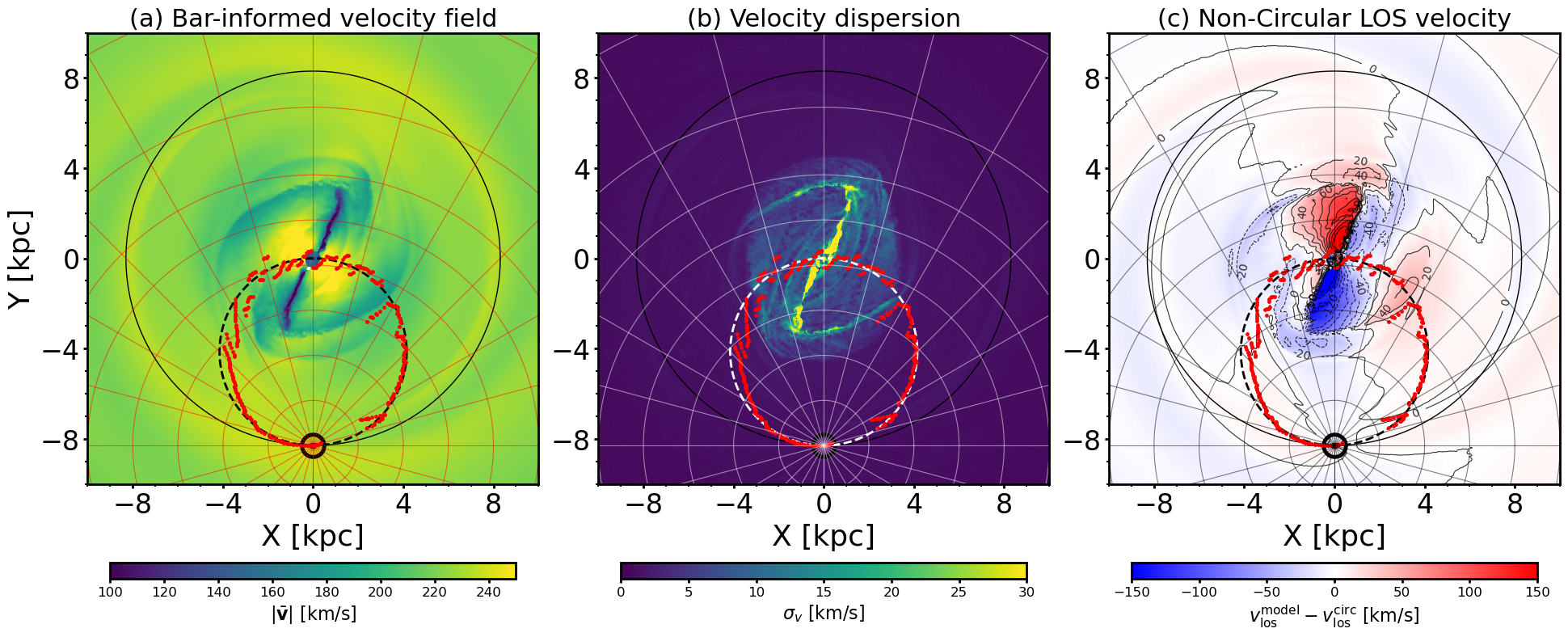}
\end{center}
\caption{
Bar-informed streaming fields used in the BIKD analysis, constructed from a quasi-steady snapshot of the hydrodynamical simulation.
\textbf{(a)} Mass-weighted mean in-plane velocity field $\bar{\mathbf{v}}(X,Y)$ in the Galactic plane, shown as the speed $|\bar{\mathbf{v}}|$ (color).
\textbf{(b)} The corresponding velocity-dispersion map $\sigma_v(X,Y)$ estimated from the local velocity scatter in the same kernel-smoothed neighborhood.
\textbf{(c)} Difference between the model and axisymmetric circular line-of-sight velocities, $\Delta v_{\rm los}(X,Y)\equiv v_{\rm los}^{\rm model}(X,Y)-v_{\rm los}^{\rm circ}(X,Y)$, as seen by an observer at $(X,Y)=(0,-R_0)$.
In all panels the Sun is marked by the $\odot$ symbol and the solid circle indicates the Solar circle ($R=R_0$).
The dashed curve shows the tangent-point locus, and the red points mark the discrete terminal points used to define terminal velocities along each longitude. Overlaid grids indicate lines of constant heliocentric distance and Galactic longitude.
\textbf{Alt text}: Three-panel maps in the Galactic plane $(X,Y)$: (a) mean in-plane streaming speed $|\bar{\mathbf{v}}|$, (b) velocity dispersion $\sigma_v$, and (c) $\Delta v_{\rm los}=v_{\rm los}^{\rm model}-v_{\rm los}^{\rm circ}$ for an observer at $(0,-R_0)$, all shown by color. 
The Sun ($\odot$), the Solar circle ($R=R_0$), the tangent-point locus (dashed), and terminal points (red) are marked; grid lines indicate constant heliocentric distance and Galactic longitude.
}
\label{fig:model_vfield_vlos}
\end{figure*}

We adopt a right-handed Galactocentric Cartesian coordinate system $(X,Y,Z)$, where the Galactic center is at the origin and the Sun is at $(X,Y,Z)=(0,-R_0,0)$.
For a line of sight $(l,b)$ and heliocentric distance $s$, the Galactocentric position is
\begin{equation}
\mathbf{x}(l,b,s)=\mathbf{x}_\odot + s\,\hat{\mathbf{r}}(l,b),
\end{equation}
with $\mathbf{x}_\odot=(0,-R_0,0)$ and the line-of-sight unit vector
\begin{equation}
\hat{\mathbf{r}}(l,b)=
\left(
-\sin l \cos b,\;
\cos l \cos b,\;
\sin b
\right),
\end{equation}
consistent with our convention where $l=0$ points toward $+Y$ and {
increasing $l$ rotates the line of sight toward $-X$ (anti-clockwise).}

We perform hydrodynamical simulations of gas flow using the {\tt ASURA} code \citep{Saitoh+2008,SaitohMakino2013} in a fixed external potential $\Phi(\mathbf{x})$ corresponding to an observationally constrained barred Milky Way model.
Specifically, we adopt the {\tt AGAMA} implementation \citep{Vasiliev2019} of the analytic fit by \citet{Sormani+2022agama} to the M2M bulge/bar model, constrained by bulge/bar star counts and stellar kinematics \citep{Portail+2017}. For the remaining components (stellar disks, dark-matter halo, and nuclear stellar cluster/disk), we assume a slightly modified version of the model in \citet{Hunter+2024}; details of our adopted composite potential are given in \citet{Baba2025b}.
In this paper, we do not impose an additional rigidly rotating stellar-spiral potential; because stellar spirals are expected to be transient and differentially rotating in $N$-body models \citep[e.g.,][]{Fujii+2011,Grand+2012b,Baba+2013}, this choice isolates the impact of bar-driven non-circular motions.

From a set of snapshots we construct smooth, bar-informed kinematic fields on a regular $(X,Y)$ grid in the Galactic midplane ($Z=0$).
Although BIKD can, in principle, be extended to three dimensions, in this paper we take a first step and focus on the Galactic midplane, which captures the dominant bar-driven non-circular motions relevant for low-latitude data.
At each grid point $\mathbf{x}$ we estimate the mass-weighted mean velocity via kernel smoothing,
\begin{equation}
\bar{\mathbf{v}}(\mathbf{x}) \equiv
\frac{\sum_i m_i\, \mathbf{v}_i\, W(|\mathbf{x}-\mathbf{x}_i|,h)}
     {\sum_i m_i\, W(|\mathbf{x}-\mathbf{x}_i|,h)},
\label{eq:vbar}
\end{equation}
where $W$ is the standard cubic-spline (M4) kernel \citep[e.g.,][]{Monaghan1992} with compact support,
\begin{equation}
W(q)\propto
\begin{cases}
1-\dfrac{3}{2}q^{2}+\dfrac{3}{4}q^{3}, & 0\le q<1,\\[2pt]
\dfrac{1}{4}(2-q)^{3}, & 1\le q<2,\\[2pt]
0, & q\ge 2,
\end{cases}
\end{equation}
with $q\equiv |\mathbf{x}-\mathbf{x}_i|/h$ and a smoothing length $h=0.10~{\rm kpc}$.
We also estimate an effective 3D velocity dispersion,
\begin{equation}
\sigma_{\rm 3D}^2(\mathbf{x}) \equiv
\sum_{d\in\{x,y,z\}}
\frac{\sum_i m_i\, (v_{i,d}-\bar{v}_d(\mathbf{x}))^2\, W(|\mathbf{x}-\mathbf{x}_i|,h)}
     {\sum_i m_i\, W(|\mathbf{x}-\mathbf{x}_i|,h)}.
\label{eq:sigma3d}
\end{equation}
In the likelihood we require the line-of-sight velocity dispersion.
As a baseline approximation we assume local isotropy and adopt
\begin{equation}
\sigma_{\rm los}^2(\mathbf{x}) \simeq \frac{1}{3}\,\sigma_{\rm 3D}^2(\mathbf{x}).
\label{eq:siglos}
\end{equation}

Given the bar-informed velocity field, the model line-of-sight velocity at $(l,b,s)$ is
\begin{equation}
v_{\rm los}^{\rm model}(l,b,s) \equiv 
\left[\bar{\mathbf{v}}(\mathbf{x}(l,b,s))-\mathbf{v}_{\rm LSR}\right]\cdot \hat{\mathbf{r}}(l,b),
\label{eq:vlos_model}
\end{equation}
where $\mathbf{v}_{\rm LSR} = (-V_0,0,0)$ is the LSR velocity at the Solar position expressed in the Galactocentric frame.

Figure~\ref{fig:model_vfield_vlos} illustrates these bar-informed streaming fields in the Galactic plane: panels (a) and (b) show the kernel-smoothed mean flow $\bar{\mathbf{v}}(\mathbf{x})$ and the associated dispersion field $\sigma_{\rm los}(\mathbf{x})$, 
while panel (c) shows the corresponding deviation of the model line-of-sight velocity from an axisymmetric circular-flow prediction, $\Delta v_{\rm los}(\mathbf{x}) \equiv v_{\rm los}^{\rm model}(\mathbf{x}) - v_{\rm los}^{\rm circ}(\mathbf{x})$,
as seen by an observer at the Solar position.
In an axisymmetric model with purely circular rotation, the terminal velocity at each longitude would arise at the tangent point and therefore lie on the tangent-point locus (black dashed curve). In the barred flow, however, the maximum $|v_{\rm los}^{\rm model}|$ along a given sightline is generally reached away from the tangent point because of strong non-circular streaming. This shift is illustrated by the locus of model terminal points (red dotted curve), which departs from the tangent-point locus \citep[][]{Chemin+2015,Davis+2026}.

For comparison, previous bar-informed deprojections \citep[e.g.,][]{Pohl+2008,MertschVittino2021,MertschPhan2023} typically adopted a gas-flow family obtained from hydrodynamical simulations in the barred potential model of \citet{Bissantz+2003}. 
In that framework, the stellar mass model is derived by non-parametrically deprojecting extinction-corrected COBE/DIRBE near-infrared surface-brightness maps into a 3D luminosity density and converting it to mass with an assumed mass-to-light ratio. 
\citet{MertschVittino2021,MertschPhan2023} also considered an \citet{Sormani+2015a}-based semi-analytic bar-orbit alternative. 
Here, we instead derive the streaming field from our own hydrodynamical simulations \citep[][]{Baba2025b} in the \citet{Portail+2017}-constrained barred Milky Way potential adopted above.

\subsection{Distance posterior}
\label{subsec:BIKD_posterior}

Our goal is to infer the heliocentric distance $s$ for each $(l,b,v)$ data in the barred inner Galaxy.
Distance inference is intrinsically non-unique: along many sightlines $v_{\rm los}^{\rm model}(l,b,s)$ is non-monotonic, and unresolved motions and model imperfections broaden the relation between $v$ and $s$.
The present method targets geometric artefacts that arise when an axisymmetric circular-rotation model is applied to the barred inner Galaxy.
{
In this work we follow the standard approach used in many kinematic CO reconstructions and adopt a fixed $X_{\rm CO}$ conversion when translating integrated CO intensity to $\Sigma_{\rm H_2}$ \citep[e.g.,][]{NakanishiSofue2016,MertschVittino2021}.
We therefore do not forward-model the CO emission with an explicit radiative-transfer calculation, although such treatments exist in the literature \citep[][]{Soding+2025}.
}

\subsubsection{Posterior}
\label{subsubsec:BIKD_posterior}

Physically, PPV emission at fixed $(l,b)$ receives contributions from gas distributed along the line of sight.
Local line broadening (e.g.\ turbulence) and the instrumental spectral response spread the emission over velocity channels \citep[e.g.][]{Wilson+RadioAstronomy2013}.
We capture this effect by treating the observed line-of-sight velocity $v$ at fixed $(l,b,s)$ as a random variable under the streaming model $\mathcal{M}$.
Specifically, $p(v\mid l,b,s,\mathcal{M})$ denotes the conditional probability density for observing $v$ from gas located at heliocentric distance $s$ along the sightline $(l,b)$, given the model-predicted mean velocity and dispersion at that location.

Accordingly, Bayes' theorem gives
\begin{equation}
P(s \mid l,b,v,\mathcal{M}) \propto \pi(s \mid l,b)\, \mathcal{L}(v \mid l,b,s,\mathcal{M}),
\label{eq:bayes_grid}
\end{equation}
where $\pi(s\mid l,b)$ is the distance prior (Section~\ref{subsubsec:BIKD_prior}).
We define the likelihood as $\mathcal{L}(v \mid l,b,s,\mathcal{M}) \equiv p(v\mid l,b,s,\mathcal{M})$, viewed as a function of $s$ for a given observed $v$; its functional form is specified in Section~\ref{subsubsec:BIKD_like}.
To evaluate the posterior numerically, we discretize distance as $s_k = (k+1)\,\Delta s$ ($k=0,\dots,N_s-1$), with $s_{\rm max}=N_s\Delta s$.
On this grid, Bayes' theorem becomes $P(s_k \mid l,b,v,\mathcal{M}) \propto \pi(s_k \mid l,b)\,\mathcal{L}(v \mid l,b,s_k,\mathcal{M})$.
We normalize explicitly as
\begin{equation}
P(s_k \mid l,b,v,\mathcal{M}) =
\frac{\pi(s_k \mid l,b)\,\mathcal{L}(v \mid l,b,s_k,\mathcal{M})}
     {\sum_{k'} \pi(s_{k'} \mid l,b)\,\mathcal{L}(v \mid l,b,s_{k'},\mathcal{M})},
\label{eq:posterior_grid}
\end{equation}
so that $\sum_k P(s_k \mid l,b,v,\mathcal{M}) = 1$.
We evaluate Eqs.~(\ref{eq:bayes_grid})--(\ref{eq:posterior_grid}) robustly using a log-sum-exp normalization.
In practice, both the likelihood and the resulting distance posterior depend on the assumed streaming model $\mathcal{M}$, which specifies the velocity and dispersion fields (e.g., pattern speed, Sun--bar angle, snapshot time, and the construction of $\bar{\mathbf{v}}$ and $\sigma_{\rm los}$). We quantify this model dependence by marginalizing over an ensemble of models (Section~\ref{subsec:model_dep}).

Interpolation of the simulation-based velocity and dispersion fields can fail outside the precomputed grid or in under-sampled regions, yielding undefined values. Unless stated otherwise, we exclude such distance grid points by setting $\ln\mathcal{L}=-\infty$ (equivalently, assigning zero posterior weight before normalization); we verified that our main results are unchanged when using a simple axisymmetric fallback model for these points.

\subsubsection{Likelihood}
\label{subsubsec:BIKD_like}
We adopt a Gaussian likelihood for the observed line-of-sight velocity $v$,
\begin{equation}
\begin{aligned}
\ln \mathcal{L}(v \mid l,b,s_k,\mathcal{M})
&= -\frac{\left[v-v_{\rm los}^{\rm model}(l,b,s_k;\mathcal{M})\right]^2}
{2\,\sigma_{\rm eff}^2(l,b,s_k;\mathcal{M})} \\
&\quad -\frac{1}{2}\ln\!\left(2\pi\sigma_{\rm eff}^2(l,b,s_k;\mathcal{M})\right),
\end{aligned}
\label{eq:loglike}
\end{equation}
where the effective dispersion is
\begin{equation}
\sigma_{\rm eff}^2(l,b,s_k;\mathcal{M}) \equiv
\sigma_{\rm los}^2\!\left(\mathbf{x}(l,b,s_k);\mathcal{M}\right) + \sigma_{\rm inst}^2 + \sigma_{\rm th}^2 .
\label{eq:sigma_eff}
\end{equation}
The term $\sigma_{\rm inst}$ represents instrumental velocity uncertainty (finite channel width and measurement noise), which we treat as a constant.
The additional term $\sigma_{\rm th}$ is a nuisance broadening parameter that absorbs unresolved small-scale motions and residual mismatches between the data and the gridded streaming model (e.g., sub-grid turbulence); it is also treated as a constant and added in quadrature.
For molecular-gas tracers the true thermal width is typically much smaller than the turbulent dispersion, and thus $\sigma_{\rm th}$ should be interpreted as a generic microturbulent (model-error) term rather than literal thermal broadening. In the applications below we adopt $\sigma_{\rm inst}=2~{\rm km\,s^{-1}}$ and $\sigma_{\rm th}=7~{\rm km\,s^{-1}}$.

\subsubsection{Prior}
\label{subsubsec:BIKD_prior}

We adopt weak distance priors to avoid injecting unnecessary structure into the reconstruction.
Our baseline family is a power-law prior,
\begin{equation}
\pi(s) \propto s^{p},
\label{eq:prior_powerlaw}
\end{equation}
where $p$ controls how much prior weight is given to large heliocentric distances along a sightline.
We consider three simple choices.
(i) $p=0$: a flat prior in distance, which assigns equal prior weight to equal intervals in $s$.
(ii) $p=1$: a mildly increasing prior, used here as a disk-motivated intermediate case.
(iii) $p=2$: a volume prior, $\pi(s)\propto s^2$, which would apply for a spatially uniform three-dimensional density field because the available volume element scales as $s^2\,ds$.
Because the Galactic gas is concentrated in a thin disk and our analysis focuses on low latitudes, the $p=2$ prior can over-emphasize large distances; we therefore treat $p=2$ as an upper-limit reference case and include $p=1$ as a more conservative alternative.

In both the simulation validation (Section~\ref{sec:validation}) and the observational application (Section~\ref{sec:obs}), we repeat the reconstruction for $p=0,1,2$ to quantify prior sensitivity and to verify that our main morphological conclusions are unchanged.

\subsubsection{Posterior sampling and soft assignment}
\label{subsubsec:BIKD_estimators}

\begin{figure}
\begin{center}
\includegraphics[width=0.45\textwidth]{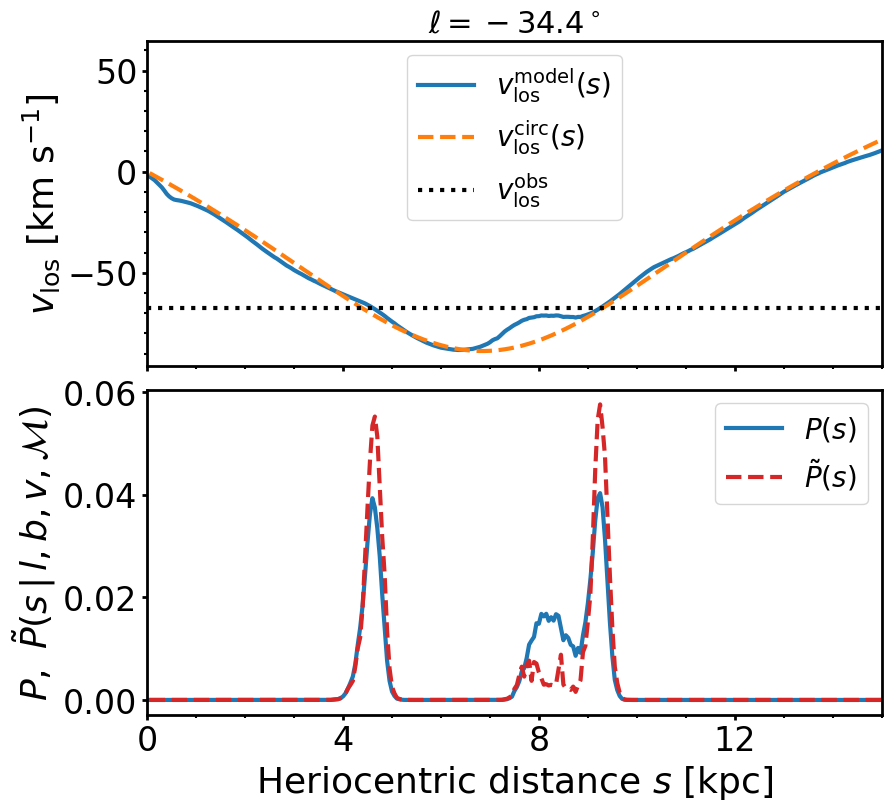}
\end{center}
\caption{
Example distance posterior for a single sightline used in the posterior-sampling map-making.
\textbf{Top}: the bar-informed model line-of-sight velocity curve $v_{\rm los}^{\rm model}(s)$ for the adopted streaming model $\mathcal{M}$ (solid), compared with the line-of-sight velocity predicted by an axisymmetric circular-rotation model $v_{\rm los}^{\rm circ}(s)$ (dashed).
The observed velocity $v_{\rm los}^{\rm obs}$ is shown as a horizontal dotted line.
\textbf{Bottom}: the corresponding normalized distance posterior $P(s|l,b,v,\mathcal{M})$ (solid) and the Jacobian-reweighted distribution $\tilde{P}(s|l,b,v,\mathcal{M})$ (dashed; eq.~(\ref{eq:posterior_for_map})) used for posterior-sampling map making.
Where the horizontal line $v_{\rm los}^{\rm obs}$ intersects the non-monotonic curve $v_{\rm los}^{\rm model}(s)$, multiple distances can be consistent with the same observed velocity, producing a multi-modal posterior (near\UTF{2013}far ambiguity).
\textbf{Alt text}: Two-panel plot for one sightline. Top: $v_{\rm los}$ versus distance $s$ showing $v_{\rm los}^{\rm model}(s)$ (solid), $v_{\rm los}^{\rm circ}(s)$ (dashed), and the observed $v_{\rm los}^{\rm obs}$ (horizontal dotted). Bottom: two curves showing $P(s|l,b,v,\mathcal{M})$ (solid) and $\tilde{P}(s|l,b,v,\mathcal{M})$ (dashed), which can be multi-modal (near\UTF{2013}far ambiguity) where $v_{\rm los}^{\rm obs}$ intersects a non-monotonic $v_{\rm los}^{\rm model}(s)$.
}
\label{fig:d_posterior}
\end{figure}

Figure~\ref{fig:d_posterior} shows a representative example of the distance inference for a single sightline. Because $v_{\rm los}^{\rm model}(s)$ can be non-monotonic, a single observed velocity can be consistent with multiple distances, producing a multi-modal posterior $P(s\mid l,b,v,\mathcal{M})$. 
We therefore adopt a soft-assignment (posterior-weighted) map-making scheme rather than using a single-point distance estimate.

In the basic soft-assignment scheme, for each longitude--latitude--velocity voxel we distribute its intensity (or mass) over distance in proportion to the discretized posterior $P(s_k\mid l,b,v,\mathcal{M})$.
In practice, we implement this posterior-weighted deposition by Monte Carlo sampling from $P(s\mid l,b,v,\mathcal{M})$: we draw distances $s$ and deposit the voxel weight at $\mathbf{x}(l,b,s)$. In the limit of many draws, this converges to the posterior-weighted (fractional) map estimate.
This estimator is equivalent to the expectation value of the map under the adopted distance distribution and naturally accounts for multi-modality (e.g., the near--far ambiguity), yielding smoother reconstructions and reducing sensitivity to multi-modal posteriors and grid/discretization effects compared to point-estimate (hard-assignment) schemes. Negative intensity values, which arise from baseline and noise fluctuations, are excluded from the deposition (i.e.\ treated as missing data) to avoid assigning unphysical negative emissivity/mass.

For comparison, one could instead adopt a point estimate such as the maximum-a-posteriori (MAP) distance, i.e.\ the distance grid point with the highest posterior probability. This corresponds to a hard-assignment scheme, in which all mass is deposited at a single location.
Our posterior-weighted (soft-assignment) approach propagates the line-of-sight distance ambiguity into the final maps and is less sensitive to multi-modality and discretization than such point-estimate (hard) assignments.

\subsubsection{Jacobian reweighting for velocity crowding}
\label{subsubsec:BIKD_jacobian}

The Bayesian posterior $P(s\mid l,b,v,\mathcal{M})$ already captures distance ambiguity for each $(l,b,v)$ data through the likelihood and the distance prior.
The issue addressed here is not the posterior itself, but the subsequent map-making step that deposits channel-integrated emission into a spatial grid.

In PPV data, emission is recorded in finite velocity channels of width $\Delta v$.
Along a monotonic branch of $v_{\rm los}^{\rm model}(s)$, a fixed $\Delta v$ corresponds to a distance interval $\Delta s \simeq \Delta v / \left|dv_{\rm los}^{\rm model}/ds\right|$.
Therefore, if one deposits emission onto a distance grid using $P(s\mid l,b,v,\mathcal{M})$ alone, regions with small $\left|dv_{\rm los}^{\rm model}/ds\right|$ (velocity crowding) can receive systematically larger contributions per velocity channel because a larger physical path length is compressed into the same velocity width.
{
We interpret this as a deposition-weighting bias (velocity crowding) arising from the non-uniform correspondence between fixed-width velocity channels and path-length intervals along the line of sight.}

To mitigate this effect in the map-making step, we reweight the discretized posterior by the local gradient of the model curve and define
\begin{equation}
\tilde P(s_k\mid l,b,v,\mathcal{M}) \propto P(s_k\mid l,b,v,\mathcal{M})\,
\left[\max\!\left(\left|\frac{d v_{\rm los}^{\rm model}}{ds}\right|,J_{\rm floor}\right)\right]^{\beta},
\label{eq:posterior_for_map}
\end{equation}
followed by renormalization over $s$ for each $(l,b,v)$. Here $J_{\rm floor}$ prevents excessive downweighting near turning points where $dv_{\rm los}^{\rm model}/ds \rightarrow 0$, and $\beta$ controls the strength of the correction; in our fiducial reconstruction we adopt $\beta=1$ and $J_{\rm floor}=2~{\rm km\,s^{-1}\,kpc^{-1}}$. 

The lower panel of Figure~\ref{fig:d_posterior} illustrates this step by comparing $P$ and $\tilde{P}$ for the same sightline: the Jacobian reweighting downweights distances where $|dv_{\rm los}^{\rm model}/ds|$ is small and correspondingly reduces the tendency to over-deposit emission into velocity-crowded regions.
{
This $\beta$-reweighting is introduced as a practical map-making regularization and is validated by the closed-loop tests in Section~\ref{sec:validation}, rather than being derived from a change of variables in an intensity-domain integral that would conserve a specific physical quantity \citep[cf.][]{Pohl+2008}. In practice, it tends to yield more compact face-on structures by reducing over-deposition in velocity-crowded regions. 
An alternative approach is to impose an explicit spatial correlation prior (regularization)  in the map domain \citep[][]{MertschVittino2021,MertschPhan2023,Soding+2025}. Implementing this would require a substantially more complex joint map inference and is beyond the scope of this paper.}

Unless stated otherwise, all posterior-sampling (soft-assignment) maps in this paper are constructed by sampling from $\tilde{P}$; if the reweighting above is disabled, then $\tilde{P}\equiv P$.

\section{Validation on hydrodynamical simulations}
\label{sec:validation}

Before applying BIKD to observational data cubes, we validate the full inference and map-making pipeline in a closed-loop experiment using the simulation itself.
We generate synthetic $(l,b,v)$ data by forward-projecting the simulation particles using the same observer convention and forward model as in Section~\ref{sec:BIKDmethod}, restricting to low latitudes $|b|<b_{\rm max}$.
We treat each SPH particle as a test particle and infer the discrete distance posterior $P(s_k\mid l,b,v,\mathcal{M})$ on the grid $\{s_k\}$, and construct the corresponding map-making distribution $\tilde{P}(s_k\mid l,b,v,\mathcal{M})$ following Section~\ref{subsubsec:BIKD_estimators}.
Distance uncertainty is propagated exclusively via posterior sampling from $\tilde{P}$.

We validate the map-making step by comparing the reconstructed face-on surface-density map $\Sigma_{\rm rec}(X,Y)$ with the corresponding true map $\Sigma_{\rm true}(X,Y)$ constructed from the same particle set using identical binning, latitude cuts, and selection.
We construct $\Sigma_{\rm rec}(X,Y)$ by drawing $K$ distance samples per particle from $\tilde{P}(s\mid l,b,v,\mathcal{M})$, converting each draw to $(X,Y)$, and depositing an equal mass fraction $m_i/K$ at the sampled positions; division by the pixel area yields the surface density. We adopt $K = 64$ samples per particle.
We quantify agreement using the log-residual
\begin{equation}
\Delta \ln \Sigma(X,Y) \equiv \ln \Sigma_{\rm rec}(X,Y) - \ln \Sigma_{\rm true}(X,Y).
\end{equation}

\subsection{Sensitivity to the distance prior}
\label{subsec:validation_prior_sens}

\begin{figure*}
\begin{center}
\includegraphics[width=1.\textwidth]{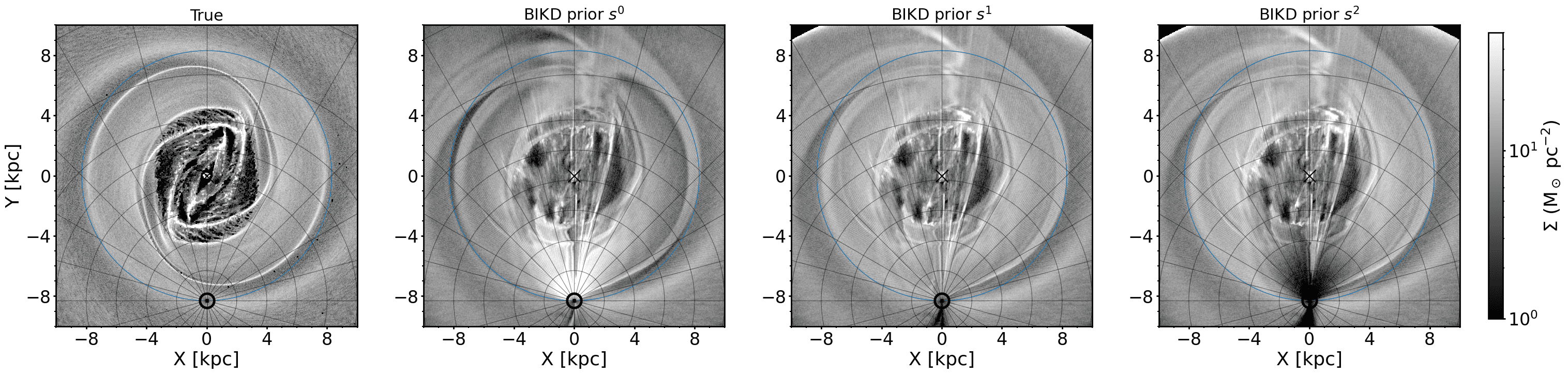}
\end{center}
\caption{
Validation of the BIKD reconstruction against the true face-on gas surface-density map in the simulation, and sensitivity to the distance prior.
From left to right, the panels show the true surface density $\Sigma_{\rm true}(X,Y)$ in the Galactic plane, followed by BIKD reconstructions $\Sigma_{\rm rec}(X,Y)$ obtained with posterior sampling and soft assignment for three power-law distance priors $\pi(s)\propto s^{p}$: $p=0$ (flat), $p=1$, and $p=2$.
\textbf{Alt text}: Four face-on maps in the Galactic plane $(X,Y)$. The left panel shows the true simulated gas surface density $\Sigma_{\rm true}(X,Y)$, and the three panels to the right show BIKD-reconstructed surface-density maps $\Sigma_{\rm rec}(X,Y)$ using posterior sampling and soft assignment with distance priors $\pi(s)\propto s^{p}$ for $p=0$ (flat), $p=1$, and $p=2$.
}
\label{fig:sim_bikd_map}
\end{figure*}

Figure~\ref{fig:sim_bikd_map} compares the true map with BIKD reconstructions obtained via posterior sampling for three power-law distance priors, $\pi(s)\propto s^{p}$, with $p=0,1,2$.
A key goal of BIKD is that the large-scale morphology is driven primarily by the kinematic likelihood (through the streaming field), rather than by strong prior assumptions.
We therefore assess prior sensitivity using (i) a visual comparison of the reconstructed face-on maps and (ii) pixel-level residual statistics between the reconstructed and true maps, summarized by the median (bias) and the median absolute deviation (MAD; scatter).

Across the three priors, the reconstructed inner-Galaxy morphology---most notably the bar-like elongation---is qualitatively stable, indicating that the streaming-informed likelihood dominates where the model is informative.
At the same time, the prior affects the inferred normalization at small heliocentric distances: the $p=0$ (flat) prior yields systematically higher inferred surface densities nearby, whereas the $p=2$ (volume) prior suppresses them.
Among the three choices, the intermediate $p=1$ prior provides the best overall match to the true surface-density distribution in this validation.
Quantitatively, the $p=1$ prior yields the smallest pixel-to-pixel scatter in $\Delta\ln\Sigma$ (MAD $=0.29$ within $R<8~\mathrm{kpc}$), while keeping the global bias small (median $=0.07$; Table~\ref{tab:prior_dlnsigma_compact}).
The $p=0$ prior shows a substantially larger scatter (MAD $=0.44$), and the $p=2$ prior exhibits a comparable bias but a larger scatter than $p=1$ (Table~\ref{tab:prior_dlnsigma_compact}).

Taken together, these tests show that the BIKD reconstruction is only weakly sensitive to the assumed distance prior at the level of large-scale inner-Galaxy morphology, while the prior primarily affects the local (near-Sun) normalization.
For the remainder of this paper, we therefore adopt the intermediate $p=1$ prior as our fiducial choice, because it provides the best overall agreement with the true map in the closed-loop validation and minimizes the residual scatter in $\Delta\ln\Sigma$ (Table~\ref{tab:prior_dlnsigma_compact}).
We emphasize that this choice is made to reduce prior-driven normalization biases, whereas the qualitative bar-scale features discussed below are robust across the tested priors.

\begin{table}
\centering
\caption{
Summary statistics of the log-residual $\Delta\ln\Sigma$ for the distance-prior test shown in Fig.~\ref{fig:sim_bikd_map}, evaluated over pixels within $R<8~\mathrm{kpc}$.
For each prior, $\Delta\ln\Sigma$ is computed between the BIKD reconstruction and the corresponding true map in Fig.~\ref{fig:sim_bikd_map}.
Note that $\Delta\ln\Sigma$ (and hence its median and MAD) is dimensionless.
}
\label{tab:prior_dlnsigma_compact}
\begin{tabular}{lcc}
\hline
Prior & $\mathrm{median}$ & $\mathrm{MAD}$ \\
\hline
$p=0$ (flat)    & $-0.0929$ & $0.4406$ \\
$p=1$           & $ 0.0723$ & $0.2922$ \\
$p=2$           & $ 0.0928$ & $0.3305$ \\
\hline
\end{tabular}
\end{table}

\subsection{Sensitivity to the streaming-field realization}
\label{subsec:model_dep}

BIKD relies on a simulation-derived, non-axisymmetric streaming model to map each $(l,b,v)$ data to a distance posterior.
We therefore test the sensitivity of the reconstruction to plausible variations in the streaming field and the observer configuration, and use these variations to define an ensemble for marginalization.

\begin{figure*}
\begin{center}
\includegraphics[width=1.\textwidth]{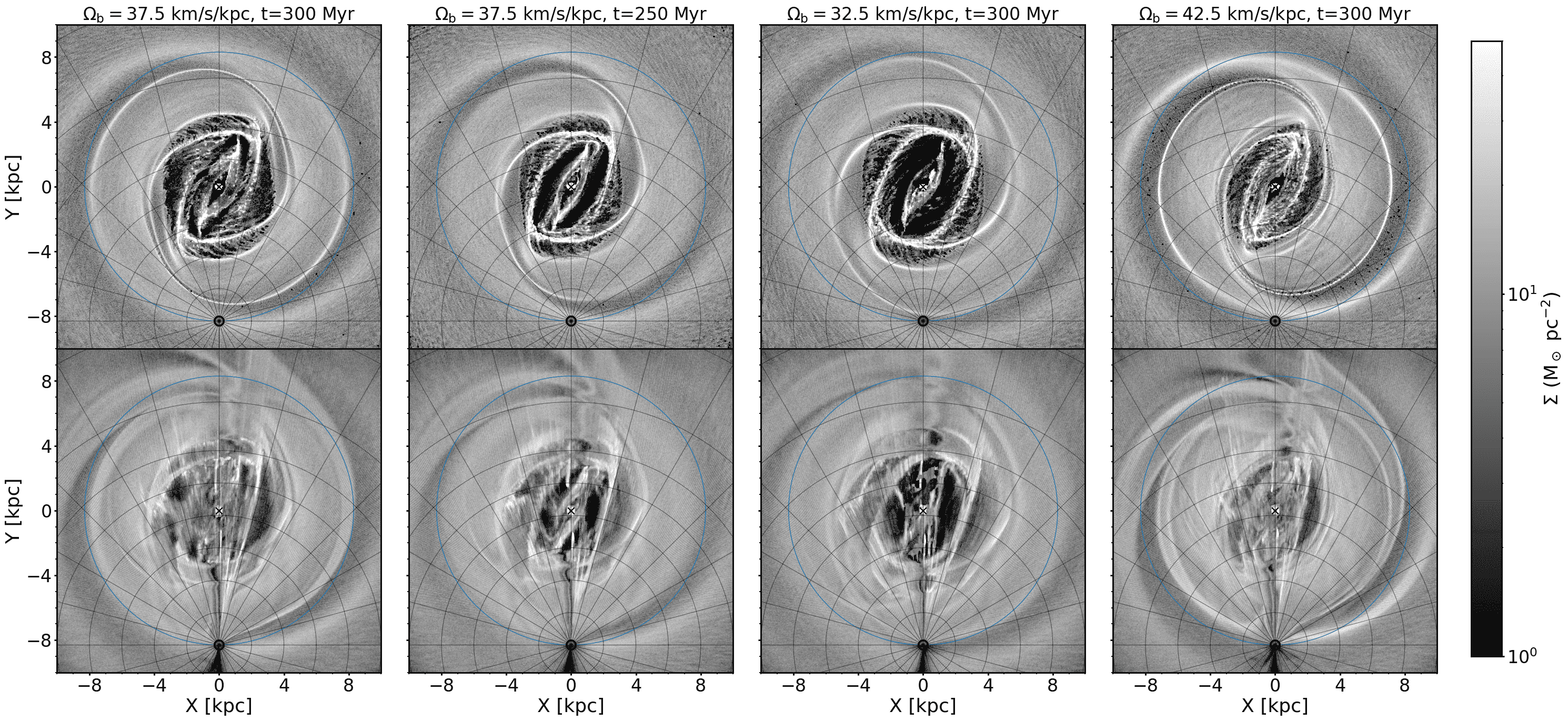}
\end{center}
\caption{
Model dependence of the BIKD reconstruction in hydrodynamical simulations.
Each column corresponds to a different streaming-field realization used by BIKD: the fiducial model with $\Omega_{\rm b}=37.5~\mathrm{km\,s^{-1}\,kpc^{-1}}$ at $t=300~\mathrm{Myr}$ (\textbf{first} column), the same pattern speed at an earlier snapshot $t=250~\mathrm{Myr}$ (\textbf{second}), and simulations with $\Omega_{\rm b}=32.5$ and $42.5~\mathrm{km\,s^{-1}\,kpc^{-1}}$ at $t=300~\mathrm{Myr}$ (\textbf{third} and \textbf{fourth}).
Top panels show the true face-on gas surface density constructed directly from the simulations.
Bottom panels show the corresponding BIKD reconstructions obtained by posterior-sampling map-making, using the streaming field derived from each realization.
\textbf{Alt text}: Eight-panel figure arranged as four columns and two rows. Columns correspond to different streaming-field realizations (fiducial $\Omega_{\rm b}=37.5~\mathrm{km\,s^{-1}\,kpc^{-1}}$ at $t=300~\mathrm{Myr}$; the same $\Omega_{\rm b}$ at $t=250~\mathrm{Myr}$; and $\Omega_{\rm b}=32.5$ and $42.5~\mathrm{km\,s^{-1}\,kpc^{-1}}$ at $t=300~\mathrm{Myr}$). The top row shows the true simulated face-on gas surface density, and the bottom row shows the corresponding BIKD-reconstructed maps for each realization.
}
\label{fig:sim_bikd_model_cmp}
\end{figure*}

Figure~\ref{fig:sim_bikd_model_cmp} illustrates the dependence on snapshot time and bar pattern speed.
Each column shows a closed-loop reconstruction: we forward-project simulation particles to synthetic $(l,b,v)$ data and then apply BIKD using a streaming field extracted from a specific simulation realization.
The top row shows the true face-on surface density for that realization, while the bottom row shows the corresponding BIKD reconstruction.
The first two columns compare two snapshots within the quasi-steady phase at fixed $\Omega_{\rm b}=37.5~\mathrm{km\,s^{-1}\,kpc^{-1}}$ (fiducial), at $t=300$ and $250~\mathrm{Myr}$.
The remaining columns vary the pattern speed at fixed snapshot time ($t=300~\mathrm{Myr}$), spanning $\Omega_{\rm b}=32.5$--$42.5~\mathrm{km\,s^{-1}\,kpc^{-1}}$, motivated by recent observational constraints \citep[e.g.][]{HuntVasiliev2025} and centered on the best-fit made-to-measure value.

Across these realizations, the main morphological features recovered by BIKD---including the bar-like elongation, the molecular ring, and the relative deficit at intermediate radii---remain qualitatively stable, indicating that the reconstruction is not driven by a single assumed streaming field.
This stability is also supported quantitatively: the pixel-level log-residual statistics $\Delta\ln\Sigma$ vary only weakly across the tested realizations, with similar medians and MADs within $R<8~\mathrm{kpc}$ (Table~\ref{tab:model_dlnsigma_compact}).
Over $R<8~\mathrm{kpc}$, the median changes by only $\simeq 0.01$ and the MAD by $\simeq 0.02$ across the tested cases, suggesting that within the explored ranges of snapshot time and $\Omega_{\rm b}$ the reconstruction error budget is not dominated by the particular choice of streaming-field realization.

Because this closed-loop test does not indicate that any single realization provides a systematically better reconstruction (Table~\ref{tab:model_dlnsigma_compact}), we adopt uniform weights in the ensemble marginalization (Section~\ref{subsec:mitigating_model_dep}), treating all sampled realizations as equally plausible within the explored ranges.

\begin{table}
\centering
\caption{
Summary statistics of the log-residual $\Delta\ln\Sigma$ for the model-dependence test shown in Fig.~\ref{fig:sim_bikd_model_cmp}, evaluated over pixels within $R<8~\mathrm{kpc}$.
For each realization, $\Delta\ln\Sigma$ is computed between the BIKD reconstruction and the corresponding true map from the same simulation snapshot.
}
\label{tab:model_dlnsigma_compact}
\begin{tabular}{lcc}
\hline
Model & median & MAD \\
\hline
$\Omega_{\rm b}=37.5$ $t=300~\mathrm{Myr}$    & $0.0723$ & $0.2922$\\
$\Omega_{\rm b}=37.5$ $t=250~\mathrm{Myr}$    & $0.0765$ & $0.2823$\\
$\Omega_{\rm b}=32.5$ $t=300~\mathrm{Myr}$    & $0.0801$ & $0.2988$\\
$\Omega_{\rm b}=42.5$ $t=300~\mathrm{Myr}$    & $0.0830$ & $0.3057$\\
\hline
\end{tabular}
\end{table}

\subsection{Ensemble marginalization and a systematic-uncertainty map}
\label{subsec:mitigating_model_dep}

In the full analysis we marginalize over an ensemble of model realizations $\mathcal{M}\equiv(\phi_{\rm bar}, t, \Omega_{\rm b})$.
We sample the Sun--bar angle at $\phi_{\rm bar}=15^\circ$, $20^\circ$, $25^\circ$, $30^\circ$, and $35^\circ$ consistent with observational constraints $\phi_{\rm bar}\approx 25^\circ\pm 10^\circ$ \citep[][and references therein]{Bland-HawthornGerhard2016,HuntVasiliev2025}.
To capture temporal fluctuations, we use six snapshots from the quasi-steady phase, namely $t=200$, $220$, $240$, $260$, $280$, and $300~\mathrm{Myr}$.
For the bar pattern speed, recent studies suggest $\Omega_{\rm b}\approx 30$--$40~\mathrm{km\,s^{-1}\,kpc^{-1}}$ \citep[][and references therein]{HuntVasiliev2025};
we therefore adopt five values centered on the best-fit made-to-measure estimate, $\Omega_{\rm b}=37.5~\mathrm{km\,s^{-1}\,kpc^{-1}}$ \citep{Portail+2017,Sormani+2022agama}, namely $\Omega_{\rm b}=32.5,\ 35.0,\ 37.5,\ 40.0$, and $42.5~\mathrm{km\,s^{-1}\,kpc^{-1}}$.
The resulting Cartesian product defines a grid of $5\times 6\times 5=150$ realizations, which we treat as equally plausible within the explored ranges.

For each realization $\mathcal{M}$ we construct a reconstructed map $\Sigma_{\rm rec}^{(\mathcal{M})}(X,Y)$ using posterior-sampling map-making.
Our fiducial reconstruction is obtained by averaging over the discrete ensemble with uniform weights,
\begin{equation}
\overline{\Sigma}(X,Y) \equiv \frac{1}{N_{\rm ens}} \sum_{\mathcal{M}} \Sigma_{\rm rec}^{(\mathcal{M})}(X,Y),
\end{equation}
where $N_{\rm ens}$ is the number of realizations in the ensemble.
This ensemble mean can be interpreted as a model-averaged BIKD map: structures that persist across realizations are reinforced, whereas features that depend sensitively on the adopted streaming field or observer configuration are suppressed.

We quantify the residual model dependence from the ensemble scatter at each pixel.
Specifically, we compute the 16th and 84th percentiles across the ensemble, $\Sigma_{16}(X,Y)$ and $\Sigma_{84}(X,Y)$, and define a fractional systematic uncertainty map as
\begin{equation}
f_{\rm sys}(X,Y)\equiv \frac{\sigma_{\rm sys}}{\overline{\Sigma}}
\equiv \frac{\left[\Sigma_{84}(X,Y)-\Sigma_{16}(X,Y)\right]/2}{\overline{\Sigma}(X,Y)},
\end{equation}
which is dimensionless.
Here $f_{\rm sys}$ measures the model-driven uncertainty associated with the finite set of plausible streaming-field realizations (not the statistical noise of the input data).

\begin{figure}
\begin{center}
\includegraphics[width=0.48\textwidth]{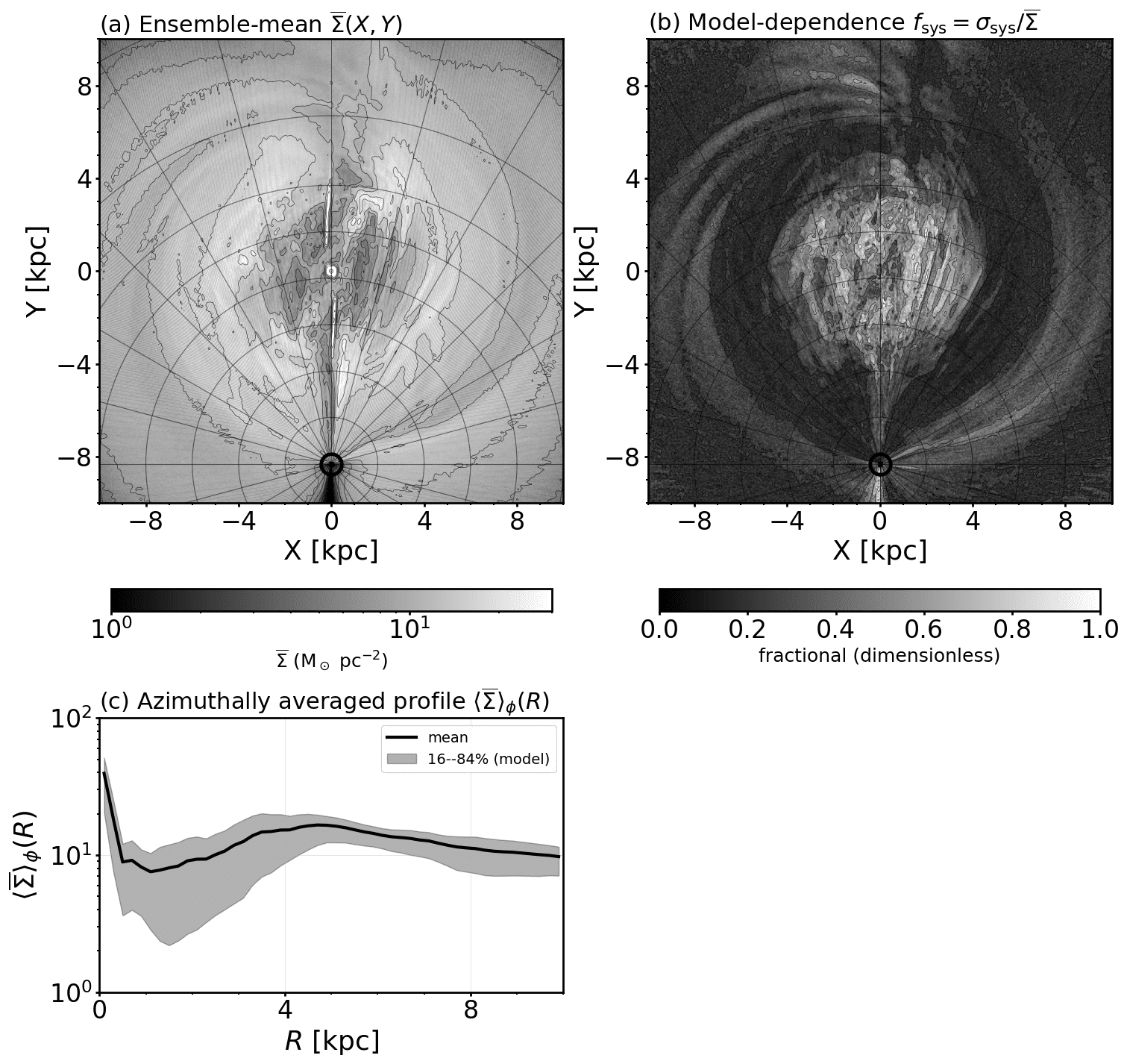}
\end{center}
\caption{
Ensemble-averaged BIKD molecular-gas surface-density map and residual model dependence in hydrodynamical simulations.
\textbf{(a)} Ensemble-mean surface density $\overline{\Sigma}(X,Y)$ obtained by averaging the BIKD-reconstructed maps over the full ensemble with uniform weights.
\textbf{(b)} Fractional model-dependence map, $f_{\rm sys}(X,Y)\equiv \sigma_{\rm sys}/\overline{\Sigma} = \big[(\Sigma_{84}-\Sigma_{16})/2\big]/\overline{\Sigma}$, where $\Sigma_{16}$ and $\Sigma_{84}$ are the per-pixel 16th and 84th percentiles across the ensemble.
\textbf{(c)} Azimuthally averaged radial profile $\langle \overline{\Sigma}\rangle_\phi(R)$; the shaded band indicates the 16--84\% range from the ensemble.
\textbf{Alt text}: Three-panel summary of ensemble-averaged BIKD results in simulations: (a) mean surface-density map $\overline{\Sigma}(X,Y)$, (b) fractional model-dependence map $f_{\rm sys}(X,Y)$ from the 16th--84th percentile spread across the ensemble, and (c) azimuthally averaged radial profile $\langle \overline{\Sigma}\rangle_\phi(R)$ with a shaded 16--84\% band.
}
\label{fig:sim_bikd_model_mean}
\end{figure}

Figures~\ref{fig:sim_bikd_model_mean}(a) and \ref{fig:sim_bikd_model_mean}(b) show the ensemble-mean map $\overline{\Sigma}(X,Y)$ and the fractional systematic uncertainty map $f_{\rm sys}(X,Y)$, respectively.
The ensemble mean highlights coherent large-scale structures that are robust to the explored model variations, including gas broadly aligned with the bar, excess emission near the bar ends, and the nuclear ring, as well as a relative deficit at intermediate radii (the bar gap).
By construction, finer substructures that appear in individual realizations (Fig.~\ref{fig:sim_bikd_model_cmp}) are partially smoothed in the ensemble mean, indicating that they are not consistently recovered across the ensemble.
The systematic-uncertainty map shows that the model dependence is largest in the barred region ($R\lesssim 4$ kpc), where $f_{\rm sys}$ typically reaches $\sim 0.6$--$0.8$; thus, in the inner Milky Way, the inferred surface density can vary at the $\sim$60--80\% level depending on the assumed realization.
This level of uncertainty is substantial, implying that moderate changes in the assumed streaming field can lead to order-unity variations in the inferred $\Sigma$ in the inner Galaxy, and that single-model reconstructions should be interpreted with caution.

We also find moderate model dependence, $f_{\rm sys}\approx 0.2$--$0.4$, at radii comparable to and beyond the Solar circle. This trend is expected if bar-driven spiral structure and the associated streaming field varies across the ensemble, in particular with $\Omega_{\rm b}$, as already suggested by the differences among the individual realizations (Fig.~\ref{fig:sim_bikd_model_cmp}).

To summarize the uncertainty in a compact form, we compute azimuthally averaged profiles. Figure~\ref{fig:sim_bikd_model_mean}(c) shows the radial profile of the ensemble mean, $\langle\overline{\Sigma}\rangle_\phi(R)$, together with the 16--84\% range obtained by applying the same averaging to $\Sigma_{16}$ and $\Sigma_{84}$. On average, the surface density is reduced at intermediate radii, $R\approx 1$--$3~\mathrm{kpc}$, supporting the presence of a bar-induced gap as a robust, ensemble-level feature. The ensemble spread indicates a model-dependence uncertainty of order $\pm 5~M_\odot\,\mathrm{pc}^{-2}$ over $R\approx 1$--$4~\mathrm{kpc}$, which we adopt as a systematic error budget when quoting amplitudes.

\subsection{Sensitivity to streaming-model mis-specification}
\label{subsec:stream_misspec}

To isolate the impact of streaming-model mis-specification, we perform an intentionally mismatched experiment.
We apply BIKD to synthetic data generated from purely circular-orbit particles (e.g., an axisymmetric initial-condition distribution), while adopting a barred, non-axisymmetric streaming field in the inference.

\begin{figure}
\begin{center}
\includegraphics[width=0.48\textwidth]{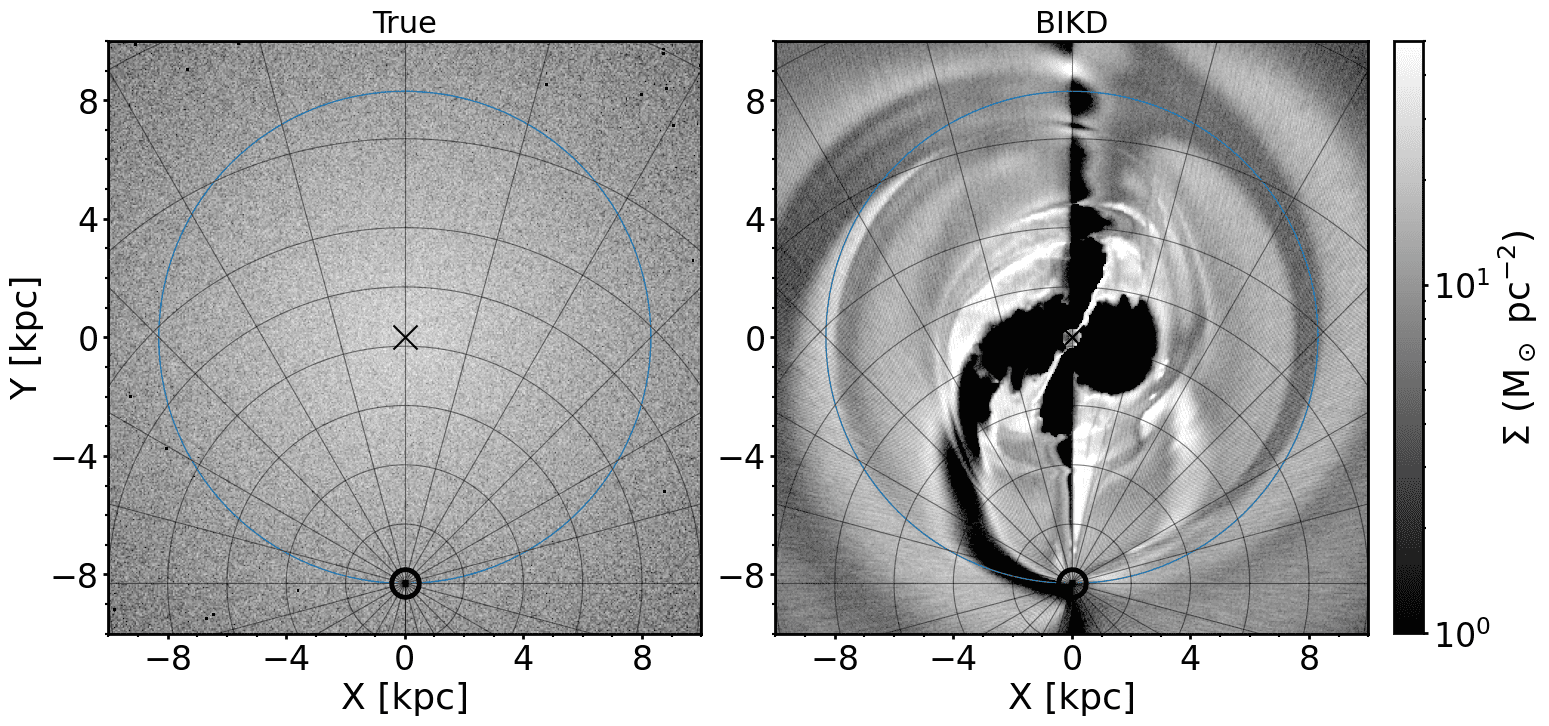}
\end{center}
\caption{
BIKD reconstruction for a circular-orbit gas distribution analyzed with a barred streaming field.
\textbf{Left:} True face-on surface-density map of the axisymmetric circular-orbit gas.
\textbf{Right:} BIKD-reconstructed map obtained by applying the bar-induced non-axisymmetric streaming model.
\textbf{Alt text}: Two face-on surface-density maps of an axisymmetric circular-orbit gas distribution. Left panel shows the true simulated map, and right panel shows the BIKD reconstruction obtained when a barred (non-axisymmetric) streaming field is used.
}
\label{fig:sim_bikd_cir}
\end{figure}

Figure~\ref{fig:sim_bikd_cir} compares the true distribution with the corresponding BIKD reconstruction.
The resulting reconstruction shows prominent artifacts at low longitudes, $|l|\lesssim 20^\circ$, and in the inner Milky Way, $R\lesssim 3~\mathrm{kpc}$, where the barred streaming field most strongly modifies the line-of-sight velocity--distance relation $v_{\rm los}^{\rm model}(s)$.
In this regime, the circular-orbit data do not populate the high-$|v_{\rm los}|$ locus predicted by the barred streaming model.
As a result, for many $(l,b,v)$ voxels the observed line-of-sight velocity lies outside (or near the edge of) the range spanned by $v_{\rm los}^{\rm model}(s)$ along the corresponding sightline.
The likelihood then provides no viable distance solution over a substantial interval in $s$, yielding negligible posterior mass at distances that project to the bar region.
When propagated through posterior-sampling map-making, this produces an apparent deficit (a ``hole'') in the inferred surface density in the bar region.

Conversely, spurious offset ridges can arise when the circular-orbit $v_{\rm los}^{\rm obs}$ coincidentally intersects a non-circular branch of $v_{\rm los}^{\rm model}(s)$.
Because $v_{\rm los}^{\rm model}(s)$ is often non-monotonic at $|l|\lesssim 20^\circ$, such accidental intersections can occur at well-defined distances, causing the posterior to place coherent probability mass at those solutions.
Posterior sampling then deposits mass at the corresponding $(X,Y)$ locations, creating false ridge-like features even though the underlying particle set is purely circular.

While \citet{Pohl+2008} noted that residual artifacts can arise when the adopted barred flow is imperfect, our intentionally mismatched test isolates this effect and shows how it propagates into deficits and false ridges in posterior-sampling map making.

This mismatch experiment demonstrates that adopting an incorrect streaming field can create both apparent deficits and false ridges; we therefore marginalize over plausible streaming-field realizations and use the ensemble scatter as a guide to model sensitivity.

\section{Application to observational data}
\label{sec:obs}

We now apply BIKD to an observed spectral-line cube.
We focus on CO emission in the inner Milky Way, where bar-driven non-circular motions strongly affect kinematic-distance reconstructions.
We use the Galactic CO($1$--$0$) survey compiled by \citet{Dame+2001}\footnote{Data available at \url{https://lweb.cfa.harvard.edu/rtdc/CO/CompositeSurveys/}.}, treating each voxel $(l,b,v)$ as carrying an intensity $I(l,b,v)$.

For each voxel $(l,b,v)$ we evaluate the distance posterior on a discrete distance grid $\{s_k\}$ following Section~\ref{subsec:BIKD_posterior}, i.e., using the likelihood defined by Eqs.~(\ref{eq:loglike})--(\ref{eq:sigma_eff}) and the distance prior in Eq.~(\ref{eq:prior_powerlaw}), with the posterior normalized such that $\sum_k P(s_k\mid l,b,v,\mathcal{M})=1$. We then redistribute the observed emission along the line of sight using these posterior weights. Specifically, we define the posterior-weighted, velocity-integrated emission assigned to distance bin $s_k$ as
\begin{equation}
w(l,b,s_k) = \sum_{v} \left[ I(l,b,v)\,\Delta v \right]\, \tilde P(s_k \mid l,b,v,\mathcal{M}),
\label{eq:w_lbs_obs}
\end{equation}
which satisfies $\sum_k w(l,b,s_k) = \sum_v I(l,b,v)\Delta v$ by construction.

To obtain a face-on map, we further integrate over Galactic latitude $b$ using a distance-dependent geometric cut corresponding to a fixed physical slab, $|z|<z_{\rm max}$ with $z_{\rm max}=0.15~{\rm kpc}$. For each distance $s_k$ we include only latitude pixels satisfying $|b|\le \arcsin(z_{\rm max}/s_k)$, and we weight the latitude sum by $\cos b\,\Delta b$. This choice is motivated by the fact that molecular gas traced by CO is strongly concentrated toward the Galactic midplane, with a characteristic vertical thickness of order $\lesssim 100$~pc in the Milky Way disk \citep[e.g.,][]{Ferriere2001,NakanishiSofue2006,Marasco+2017}. Adopting $|z|<z_{\rm max}$ therefore captures most of the CO-bright molecular layer while suppressing contamination from higher latitudes.

Finally, we map each $(l,b,s_k)$ element to Galactic-plane coordinates $(X,Y)$ using the observer position $(0,-R_0)$ and deposit its contribution onto a Cartesian grid. We account for the Jacobian of the transformation from $(l,s)$ to $(X,Y)$ by accumulating, alongside the deposited mass, the corresponding physical area element associated with each deposit (schematically $\Delta A_{XY}\propto s_k\,\Delta l\,\Delta s$). We then compute the face-on surface density as the ratio of the accumulated mass to the accumulated covered area in each cell. 
When converting CO intensity to an H$_2$ column density (and hence H$_2$ surface density), we adopt a constant CO-to-H$_2$ conversion factor $X_{\rm CO}=2\times10^{20}\ {\rm cm^{-2}\,(K\,km\,s^{-1})^{-1}}$.

\subsection{BIKD reconstructed maps}
\label{subsec:obs_model_dep}

Our distance inference relies on a non-axisymmetric streaming model $\mathcal{M}$ to predict the line-of-sight velocity curve $v_{\rm los}^{\rm model}(s)$ and an effective velocity dispersion along each sightline.
We therefore evaluate the sensitivity of the reconstructed molecular-gas map to plausible variations in the streaming field and in the assumed observer configuration.
Specifically, we repeat the reconstruction for barred-flow models with different bar pattern speeds $\Omega_{\rm b}$, for small changes in the bar viewing angle $\phi_{\rm b}$, and for nearby simulation snapshots within the quasi-steady phase.

\begin{figure*}
\begin{center}
\includegraphics[width=1.\textwidth]{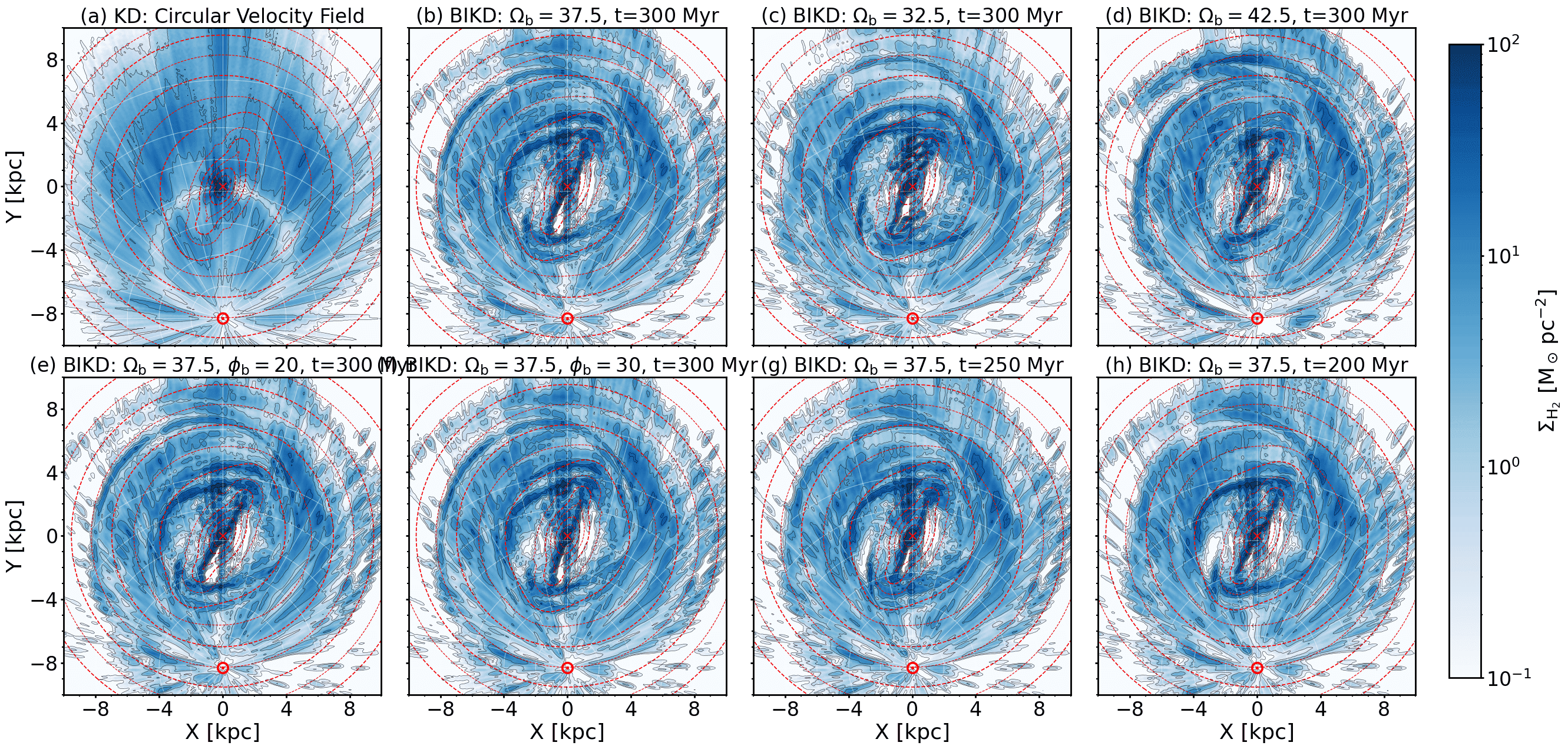}
\end{center}
\caption{
Face-on molecular-gas surface-density maps inferred from the observed Galactic CO cube using our BIKD framework, shown for systematic variations in the adopted streaming model and observer configuration.
\textbf{(a)} The reference map obtained with an axisymmetric circular-velocity field (standard KD assumption).
\textbf{(b)--(d)} BIKD reconstructions for barred-flow models that differ only in bar pattern speed, $\Omega_{\rm b}=37.5$, 32.5, and $42.5~{\rm km\,s^{-1}\,kpc^{-1}}$, respectively.
\textbf{(e,f)} The same barred-flow model with different assumed bar viewing angles: $\phi_{\rm b}=20^\circ$ and $30^\circ$ (fiducial: $25^\circ$).
\textbf{(g,h)} The same barred-flow model evaluated at earlier times (as indicated in the panel titles).
Red dashed contours trace the stellar surface-density model used to define the barred potential.
\textbf{Alt text}: Eight-panel face-on molecular-gas surface-density maps inferred from the observed CO data. (a) Standard kinematic-distance map using an axisymmetric circular-velocity field. (b--h) BIKD maps using barred streaming models with variations in pattern speed ($\Omega_{\rm b}=37.5$, $32.5$, and $42.5~\mathrm{km\,s^{-1}\,kpc^{-1}}$), bar viewing angle ($\phi_{\rm b}=20^\circ$, $25^\circ$, $30^\circ$), and snapshot time. Red dashed contours overlay the stellar surface-density model defining the barred potential.
}
\label{fig:obs_bikd_model_cmp}
\end{figure*}

Figure~\ref{fig:obs_bikd_model_cmp} summarizes these systematic tests.
The reference reconstruction based on an axisymmetric circular-velocity field (panel~a; the standard KD assumption) shows prominent line-of-sight--elongated ``finger-of-God'' (FoG) features, similar to those seen in previous KD-based maps \citep[e.g.,][]{NakanishiSofue2003,NakanishiSofue2006,NakanishiSofue2016}.
In this standard KD case, a central bar-related gas distribution is not clearly recovered.
In contrast, the BIKD reconstructions using barred-flow models (panels~b--h) substantially suppress FoG artifacts and recover a barred, quadrant-asymmetric morphology across the explored variations in $\Omega_{\rm b}$, $\phi_{\rm b}$, and snapshot time.
We quantify this suppression with an elongation statistic,
$E\equiv|\partial_\perp \ln\Sigma_{\rm H_2}|/|\partial_{\rm los}\ln\Sigma_{\rm H_2}|$,
where $\partial_{\rm los}$ and $\partial_\perp$ are derivatives along and perpendicular to the heliocentric line of sight in the $(X,Y)$ plane.
The KD map is substantially more LOS-elongated than the BIKD maps: the median $E$ is 3.6 for KD versus 1.1--1.6 for BIKD, and the fraction of pixels with $E>2$ decreases from 0.67 to 0.31--0.43.

Nevertheless, the reconstruction remains least reliable toward $|l|\lesssim 5^\circ$ even with BIKD.
Toward the Galactic center line of sight, small changes in $v_{\rm los}$ can correspond to large changes in distance because $v_{\rm los}^{\rm model}(s)$ can be non-monotonic (so the inversion is non-unique) and may vary only weakly with $s$ over extended intervals, making both KD and BIKD reconstructions susceptible to systematic distance mis-assignment and residual FoG-like stretching.
Accordingly, we do not interpret the $l\approx 0^\circ$ line-of-sight--elongated structure as a robust tracer of the true CMZ/nuclear-ring morphology.

In the standard KD case, FoG features are expected when gas with strong non-circular motions is mapped through an axisymmetric circular-velocity model \citep[e.g.,][]{Gomez2006,Baba+2009,Hunter+2024,Baba2026a}. 
In this situation, the $(l,b,v)\!\rightarrow\!s$ inversion may select the wrong distance branch and/or spread emission over a wide distance interval when $|dv_{\rm los}^{\rm circ}/ds|$ is small, making physically localized structures appear artificially elongated along the line of sight.
While these tests show that the main morphological signatures are not confined to a single choice of $\mathcal{M}$, they also illustrate that finer contrast and substructure can vary with the adopted configuration, motivating an explicit treatment of model dependence.

\begin{figure*}
\begin{center}
\includegraphics[width=1.0\textwidth]{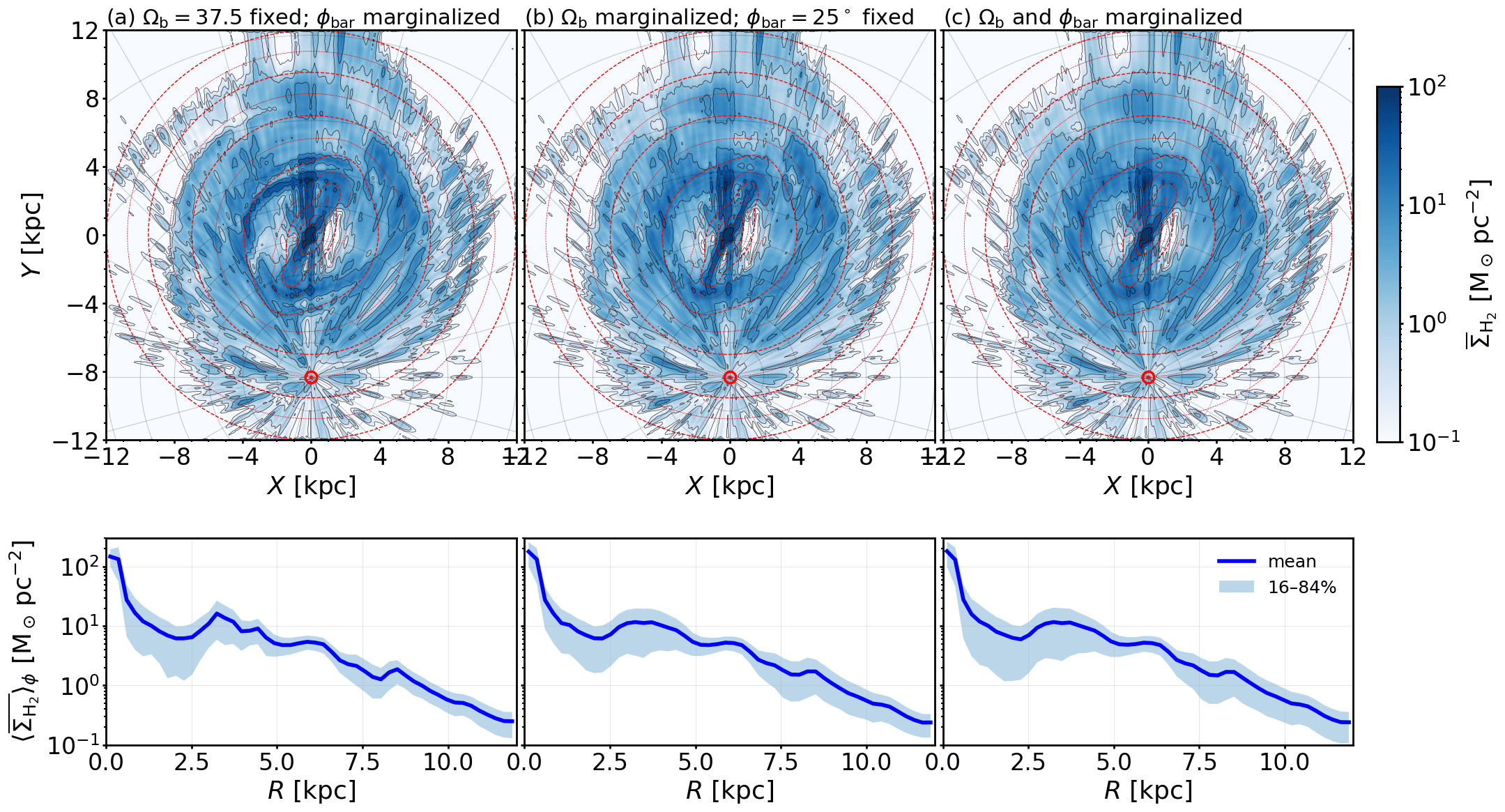}
\end{center}
\caption{
Model-marginalized BIKD reconstructions for the observed CO($1$--$0$) data under three choices of model averaging.
In all cases, we marginalize over nearby quasi-steady snapshots from $t=200$ to $300~{\rm Myr}$ in steps of $20~{\rm Myr}$.
\textbf{(a)} Fixed $\Omega_{\rm b}=37.5~{\rm km\,s^{-1}\,kpc^{-1}}$ with marginalization over the bar viewing angle $\phi_{\rm b}$.
\textbf{(b)} Marginalization over $\Omega_{\rm b}$ at fixed $\phi_{\rm b}=25^\circ$.
\textbf{(c)} Marginalization over both $\Omega_{\rm b}$ and $\phi_{\rm b}$.
\textbf{Top:} Ensemble-mean molecular-gas surface density $\overline{\Sigma}_{\rm H_2}(X,Y)$ in the Galactic plane (color scale, in ${\rm M_\odot\,pc^{-2}}$); dashed red curves show the projected stellar surface-density structure of the adopted barred model.
\textbf{Bottom:} Corresponding azimuthally averaged radial profiles $\langle \overline{\Sigma}_{\rm H_2}\rangle_\phi$.
The solid curve shows the ensemble mean, and the shaded region indicates the 16th--84th percentile range across the model ensemble.
\textbf{Alt text}: Three columns of model-marginalized BIKD results for the observed CO(1--0) data. Top row shows face-on ensemble-mean $\overline{\Sigma}_{\rm H_2}(X,Y)$ (color, $\mathrm{M_\odot\,pc^{-2}}$) with dashed red curves tracing the projected stellar structure; bottom row shows the corresponding azimuthally averaged radial profiles (mean solid, 16th--84th percentile shaded). Columns (a--c) use different marginalizations over $\Omega_{\rm b}$ and/or $\phi_{\rm b}$, including snapshots from $t=200$ to $300~\mathrm{Myr}$.
}
\label{fig:obs_bikd_model_mean}
\end{figure*}

To incorporate model dependence in a compact way, we construct model-marginalized maps by averaging over an ensemble of plausible configurations.
Figure~\ref{fig:obs_bikd_model_mean} shows three choices of model averaging: marginalization over (a) $\phi_{\rm b}$ at fixed $\Omega_{\rm b}$, (b) $\Omega_{\rm b}$ at fixed $\phi_{\rm b}$, and (c) both parameters, in all cases also marginalizing over nearby quasi-steady snapshots from $t=200$ to $300~{\rm Myr}$ in steps of $20~{\rm Myr}$.
The top panels show that all three model-marginalized BIKD maps exhibit a clear non-axisymmetric structure in the inner Milky Way ($R \lesssim 4~{\rm kpc}$) that is broadly aligned with the adopted stellar bar model \citep[red dashed contours;][]{Portail+2017,Sormani+2022agama}.
They reveal an elongated, bar-aligned enhancement in the central few kiloparsecs and pronounced asymmetries between Galactic quadrants.
When we marginalize over the bar viewing angle (panel~a), the offset ridge running roughly along the bar becomes broader, consistent with the expected smearing from small changes in $\phi_{\rm b}$.
Overall, the persistence of these bar-related features across the ensemble indicates that the large-scale CO morphology inferred by BIKD is robust to the plausible model variations considered here.

In all three cases, we also identify an inner-ring-like feature near the bar end ($R \simeq 3.5~{\rm kpc}$), as well as several spiral-arm--like, ridge-shaped overdensities outside the bar region. We compare their locations to published spiral-arm loci in Section~\ref{sec:spiral_loci}. We caution, however, that finer substructure and the detailed arm--bar transition are less well constrained, as they can depend on the adopted streaming field, the observer configuration, and the intrinsic limitations of mapping PPV data into physical space.

The bottom panels of Figure~\ref{fig:obs_bikd_model_mean} show the corresponding azimuthally averaged radial profiles of the CO-inferred molecular surface density.
All three model-marginalized reconstructions display an approximately exponential decline at $R \gtrsim 4~{\rm kpc}$, a pronounced dip at $0.5 \lesssim R \lesssim 3.5~{\rm kpc}$ (the bar gap), and a central concentration at $R \lesssim 1~{\rm kpc}$ consistent with a CMZ-level surface density of a few $\times 10^2~{\rm M_\odot\,pc^{-2}}$.
The outer boundary of this dip at $R\approx 3.5~{\rm kpc}$ is consistent with the dynamical estimate of the Galactic bar half-length \citep{Lucey+2023}.
These qualitative trends are broadly consistent with CO-based Milky Way reconstructions in the literature \citep[e.g.,][]{NakanishiSofue2006,MertschVittino2021,Soding+2025} and with generic outcomes of barred-galaxy gas-flow simulations, in which bar-driven shocks and torques concentrate gas toward the center while depleting intermediate radii \citep[e.g.][]{Athanassoula1992b,BabaKawata2020a,Sormani+2020}.
Similar radial depressions (``dips'') have been reported in a small subset of external barred spirals, suggesting that the Milky Way may be typical within that population even if such systems are uncommon in nearby-galaxy samples \citep{EvansII+2026}.

\subsection{Comparison with literature spiral-arm loci}
\label{sec:spiral_loci}

\begin{figure*}
\begin{center}
\includegraphics[height=0.9\textheight]{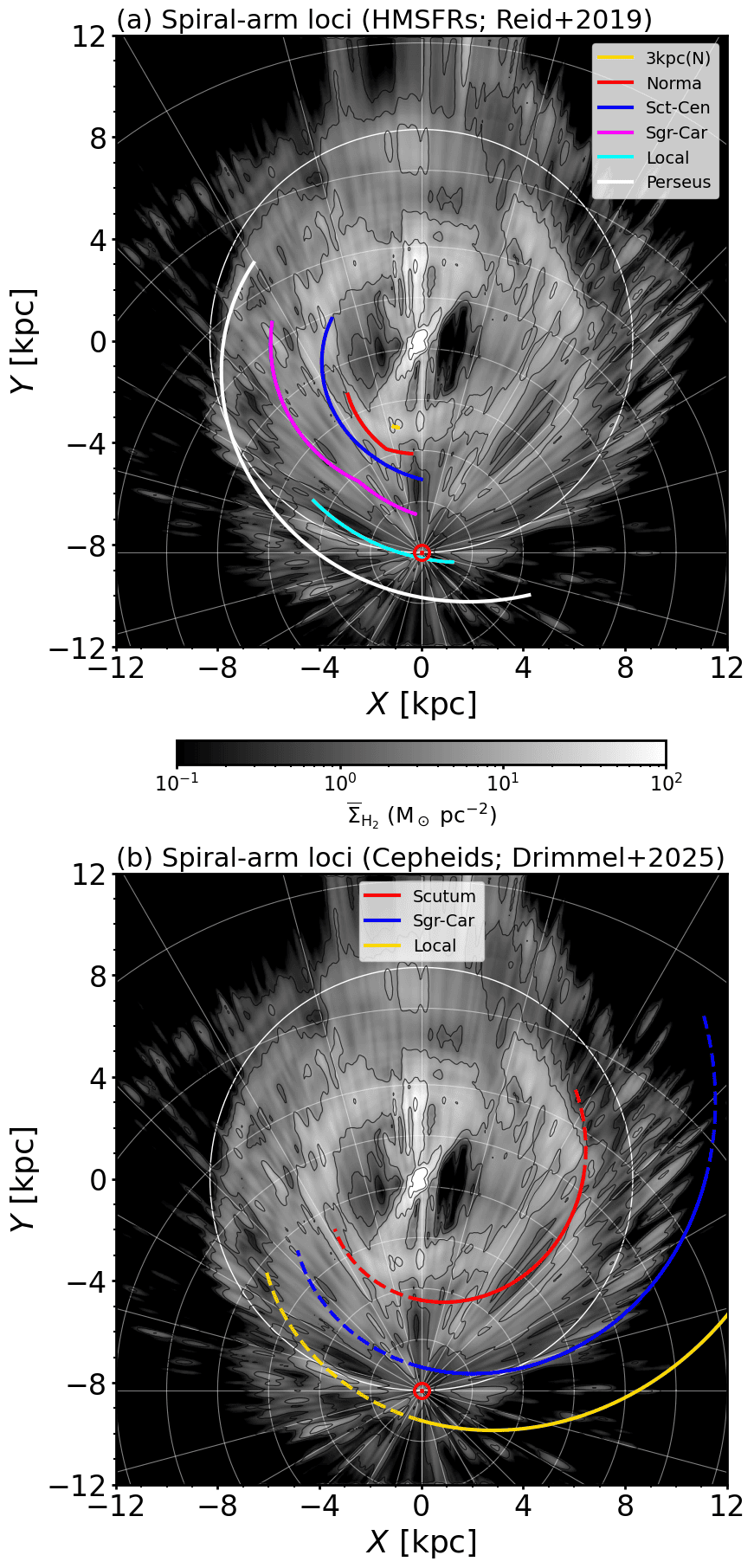}
\end{center}
\caption{
Ensemble-mean BIKD reconstructed map for the observed CO($1$--$0$) data, overlaid with literature spiral-arm loci.
\textbf{(a)} HMSFR maser-parallax loci from \citet{Reid+2019}.
\textbf{(b)} Classical-Cepheid loci from \citet{Drimmel+2025}; solid segments indicate the recommended azimuth range and dashed extensions show loci outside this range.
\textbf{Alt text}: Two-panel face-on ensemble-mean BIKD molecular-gas map from the observed CO(1--0) data with overlaid spiral-arm loci. (a) HMSFR maser-parallax arm segments from \citet{Reid+2019}. (b) Classical-Cepheid arm loci from \citet{Drimmel+2025}, with solid segments in the recommended azimuth range and dashed extensions outside it.
}
\label{fig:obs_fixWb_varAng}
\end{figure*}

We assess whether the dominant ridge-like overdensities in the BIKD molecular-gas map (from CO) occur at plausible spiral-arm phases by comparing the reconstructed molecular-gas surface-density field to independent spiral-arm loci from the literature.
We use two young-population tracers whose distance estimates are largely independent of gas kinematics: high-mass star-forming region (HMSFR) masers with VLBI trigonometric parallaxes \citep{Reid+2019} and classical Cepheids, which trace a young (tens to a few hundred Myr) stellar population and have distances primarily from mid-infrared period--luminosity relations (WISE $W1$) with extinction corrections, validated against Gaia astrometry \citep{Drimmel+2025}.
Figure~\ref{fig:obs_fixWb_varAng} overlays these loci on the BIKD map as an external, kinematics-independent consistency check; we therefore interpret the comparison statistically rather than expecting one-to-one overlap with CO ridges.

\begin{figure}
\begin{center}
\includegraphics[width=0.48\textwidth]{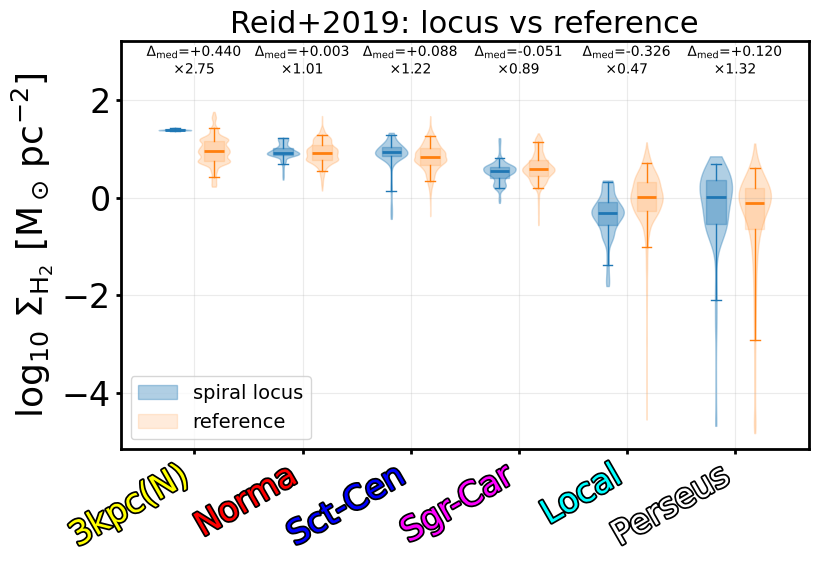}
\includegraphics[width=0.48\textwidth]{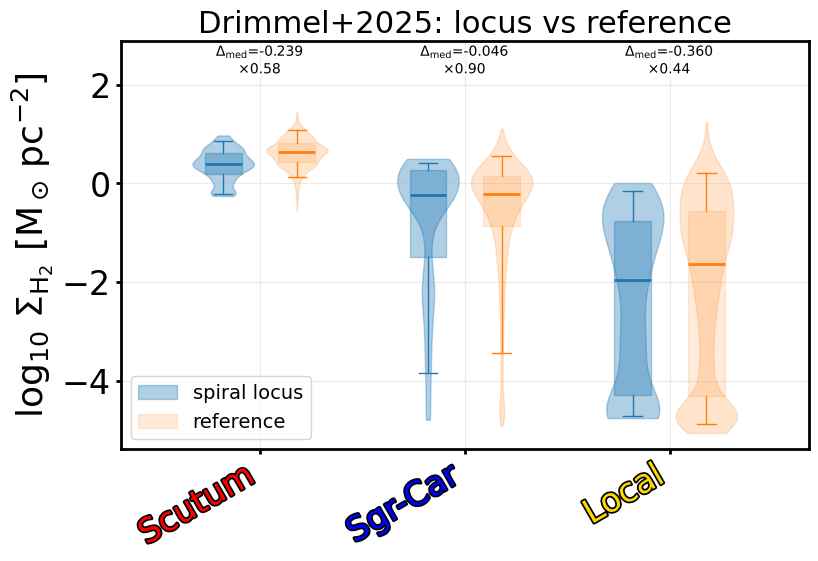}
\end{center}
\caption{
Distributions of sampled $\log_{10}\Sigma_{\rm H_2}$ along the spiral-arm loci from \citet{Reid+2019} (\textbf{upper}) and \citet{Drimmel+2025} (\textbf{lower}), compared to a matched-radial, azimuth-random reference distribution.
For each arm, the violin shows the full distribution, and the overlaid box shows the median and interquartile range (whiskers: 5th--95th percentiles).
An upward (downward) shift of the locus distribution relative to the reference indicates a higher (lower) typical CO surface density along the literature locus than expected at the same Galactocentric radius.
\textbf{Alt text}: Two stacked panels of violin-and-box plots showing the distributions of sampled $\log_{10}\Sigma_{\rm H_2}$ along spiral-arm loci, for \citet{Reid+2019} (top) and \citet{Drimmel+2025} (bottom), compared with a matched-radial azimuth-random reference. For each arm, violins show the full distributions and overlaid box plots show the median and interquartile range with 5th--95th percentile whiskers.
}
\label{fig:loci_sampling_distributions}
\end{figure}

Accordingly, rather than extracting spiral ridges from the gas map, we quantify the alignment with a ridge-free, matched-radius point-sampling test.
For each published spiral-arm locus we sample the BIKD map at the locus positions and collect the resulting values of $\Sigma_{\rm H_2}$ (the ``locus sample''). 
As a matched-radial reference, we generate comparison points by keeping the Galactocentric radius of each locus point fixed but randomizing its azimuth uniformly on the same circle, and we sample $\Sigma_{\rm H_2}$ at these randomized positions (the ``reference sample'').
This construction tests whether a literature locus preferentially traces high molecular surface density beyond what is expected from the radial surface-density trend alone.
Figure~\ref{fig:loci_sampling_distributions} compares the resulting $\log_{10}\Sigma_{\rm H_2}$ distributions for the locus sample (solid) and the matched-radial reference sample (dashed); for several loci, the locus-sample distribution is shifted toward higher values, indicating that those loci tend to intersect enhanced molecular surface density at the same radius.

To quantify the difference, we define the median offset
$\Delta_{\rm med} \equiv {\rm med}\!\left[\log_{10}\Sigma_{\rm locus}\right]-{\rm med}\!\left[\log_{10}\Sigma_{\rm ref}\right]$,
so that $\Delta_{\rm med}>0$ ($<0$) indicates that the locus typically samples higher (lower) molecular surface density than the matched-radial reference.
For convenience, we also report the corresponding multiplicative factor $10^{\Delta_{\rm med}}$.
As a complementary diagnostic, we list the two-sided Kolmogorov--Smirnov (KS) $p$-value.
Because our point sampling can be dense and neighboring samples are not strictly independent, we use $p_{\rm KS}$ mainly as evidence for a distributional difference, while $\Delta_{\rm med}$ provides the typical (median) shift.

Table~\ref{tab:loci_point_sampling_reid2019} summarizes the ridge-free point-sampling results for the \citet{Reid+2019} loci.
The 3-kpc(N) segment shows the largest median enhancement, $\Delta_{\rm med}=0.437$ ($10^{\Delta_{\rm med}}\simeq 2.74$); however, because it covers only a short longitude interval in the \citet{Reid+2019} compilation, we regard this enhancement as suggestive rather than definitive.
The Sct--Cen and Perseus loci exhibit more modest positive offsets ($10^{\Delta_{\rm med}}\simeq 1.23$ and $1.30$), indicating that these loci tend to intersect elevated molecular surface density relative to the matched-radial reference.
In contrast, the Sgr--Car and Local-arm loci yield negative offsets ($10^{\Delta_{\rm med}}\simeq 0.89$ and $0.48$), implying that these particular locus definitions do not systematically trace the dominant molecular ridges in our BIKD map.
For the Norma arm, the median offset is consistent with zero ($\Delta_{\rm med}=0.004$), i.e.\ a negligible shift in the typical surface density; the very small $p_{\rm KS}$ nonetheless indicates a detectable difference in distribution shape.

Applying the same test to the classical-Cepheid loci of \citet{Drimmel+2025} (Table~\ref{tab:loci_point_sampling_drimmel2025}), we obtain negative median offsets for all three arms considered (Scutum, Sgr--Car, and Local), with the largest deficit for Local ($\Delta_{\rm med}=-0.360$; $10^{\Delta_{\rm med}}\simeq 0.44$). Within the recommended azimuth range, these young-stellar loci therefore tend to sample lower molecular surface density than the matched-radial reference.
Taken together with the HMSFR-maser loci results, this yields a mixed picture: the inferred offsets vary in both sign and magnitude across tracers and arm segments.
Because the \citet{Reid+2019} and \citet{Drimmel+2025} loci are constrained over different azimuth ranges, the differing signs of $\Delta_{\rm med}$ should be interpreted within each tracer's azimuth coverage. A dedicated analysis with more uniform spatial coverage will be needed to assess the origin of this difference.
The mismatch likely reflects both tracer differences (molecular gas versus young stars) and the limited, non-uniform spatial coverage of the published loci, rather than a single identifiable cause.
We therefore do not interpret the loci test as validation of any specific spiral-arm model; instead, it provides a qualitative, coverage-dependent consistency check against the published loci.

Finally, we note that our streaming fields are generated from hydrodynamical simulations in a barred potential without an explicit stellar spiral potential, so any spiral structure in the model is primarily bar-driven \citep[e.g.,][]{Schwarz1981}.
Near the bar end, a bar-driven spiral response may dominate the gas morphology, so the inner-disk spiral phases in our model may still be realistic even without a prescribed stellar spiral potential \citep[e.g.,][]{Baba2015c}.
Well outside the solar circle ($R \gtrsim 10~\mathrm{kpc}$), spiral structure in the real Milky Way may be influenced by additional processes, such as tidally induced spirals driven by perturbations from the Sagittarius dwarf galaxy \citep[e.g.,][]{Antoja+2022,Asano+2025}.
Because these effects are not captured by our model, any mismatches in the outer disk should not be over-interpreted.

\begin{table}
\centering
\caption{Comparison between the BIKD CO surface-density map and the spiral-arm loci of \citet{Reid+2019}.
For each arm locus, we sample $\Sigma_{\rm H_2}(X,Y)$ along the locus and compare it to a matched-radial, azimuth-random reference sample constructed by drawing the azimuth uniformly at fixed Galactocentric radius.
We report the median offset $\Delta_{\rm med}={\rm med}[\log_{10}\Sigma_{\rm locus}]-{\rm med}[\log_{10}\Sigma_{\rm ref}]$ and the corresponding multiplicative factor $10^{\Delta_{\rm med}}$.
We also list the two-sided Kolmogorov--Smirnov (KS) $p$-value for the difference between the two distributions.}
\label{tab:loci_point_sampling_reid2019}
\begin{tabular}{lccc}
\hline
Arm & $\Delta_{\rm med}$ (dex) & $10^{\Delta_{\rm med}}$ & $p_{\rm KS}$ (two-sided) \\
\hline
3-kpc(N)     & $+0.437$ & $2.74$ & $<10^{-300}$ \\
Norma        & $+0.004$ & $1.01$ & $1.50\times 10^{-47}$ \\
Sct--Cen     & $+0.090$ & $1.23$ & $1.28\times 10^{-67}$ \\
Sgr--Car     & $-0.051$ & $0.89$ & $7.12\times 10^{-30}$ \\
Local        & $-0.316$ & $0.48$ & $4.91\times 10^{-90}$ \\
Perseus      & $+0.114$ & $1.30$ & $4.74\times 10^{-8}$  \\
\hline
\end{tabular}
\end{table}

\begin{table}
\centering
\caption{
Same as Table~\ref{tab:loci_point_sampling_reid2019}, but using the classical-Cepheid spiral-arm loci of \citet{Drimmel+2025} within their recommended azimuth range.
}
\label{tab:loci_point_sampling_drimmel2025}
\begin{tabular}{lccc}
\hline
Arm & $\Delta_{\rm med}$ (dex) & $10^{\Delta_{\rm med}}$ & $p_{\rm KS}$ (two-sided) \\
\hline
Scutum      & $-0.239$ & $0.58$ & $4.93\times 10^{-75}$ \\
Sgr--Car    & $-0.046$ & $0.90$ & $8.59\times 10^{-9}$  \\
Local       & $-0.360$ & $0.44$ & $9.99\times 10^{-6}$  \\
\hline
\end{tabular}
\end{table}

\section{Discussion and Summary}
\label{sec:summary}

We developed a bar-informed kinematic-distance framework (BIKD) to reconstruct face-on molecular-gas maps in the inner Milky Way from PPV data, replacing the standard axisymmetric circular-rotation assumption that is strongly violated in barred regions \citep[][]{Hunter+2024,Baba2026a}.
BIKD uses a non-axisymmetric streaming field extracted from hydrodynamical simulations in an observationally constrained barred potential \citep[][]{Portail+2017,Sormani+2022agama}, and infers a discrete distance posterior along each sightline with a Gaussian likelihood.
Because the resulting $v_{\rm los}^{\rm model}(s)$ relation is generically non-monotonic, we adopt posterior-weighted (soft-assignment) map making implemented via posterior sampling to propagate multi-modal posteriors into the final maps rather than relying on a single point estimate.

We validated the full pipeline in closed-loop tests on the simulations.
The recovered large-scale morphology is only weakly sensitive to simple distance priors and remains stable across plausible variations in the streaming-field realization (bar angle, snapshot time, and pattern speed).
We therefore interpret the reconstruction in a model-marginalized manner, using the ensemble scatter as a practical systematic-uncertainty indicator. The mis-specification experiment also cautions that substantially mis-specified streaming fields can introduce both apparent deficits and spurious ridges \citep[e.g.][]{Pohl+2008}.

We apply BIKD to the Dame et al.\ CO($1$--$0$) survey and obtain a face-on molecular-gas map.
Compared to a standard axisymmetric KD reconstruction, BIKD strongly suppresses the familiar line-of-sight--elongated artifacts and robustly recovers a bar-aligned, quadrant-asymmetric inner-Galaxy morphology.
The model-marginalized profiles highlight characteristic barred-flow structures, including a central concentration, an inner ring-like enhancement at $R\simeq 3$--4~kpc, and a relative deficit at intermediate radii (the bar gap) that can be obscured or distorted in axisymmetric KD maps.

We further compared ridge-shaped overdensities in the BIKD CO map with independent spiral-arm loci traced by HMSFR masers with VLBI parallaxes \citep[][]{Reid+2019} and by classical Cepheids \citep[][]{Drimmel+2025}.
Several maser-locus segments show qualitative alignment with prominent BIKD ridges, whereas other loci show little correspondence or even an apparent anti-correlation with the dominant ridges. 
The Cepheid loci likewise do not coincide with the dominant ridges within their recommended azimuthal range, and in some segments they preferentially sample inter-arm regions.
These comparisons provide an external, kinematics-independent consistency check.
A one-to-one match is not expected given tracer selection, limited azimuth coverage, and the fact that our streaming fields are generated in a barred potential without an explicit stellar spiral perturbation; thus, outer-disk spiral-arm comparisons should be interpreted with caution.

Finally, we highlight limitations and complementary directions. Both standard KD and BIKD share an intrinsic limitation: they map observed $v_{\rm los}$ to distance. Consequently, the reconstruction depends on the assumed streaming field and cannot independently constrain non-circular motions within the KD framework. This motivates complementary methods that incorporate independent distance anchors. 
Kinetic tomography \citep[][]{TchernyshyovPeek2017,Tchernyshyov+2018,Soler+2025} combines 3D dust extinction/reddening maps with H\,\textsc{i}/CO PPV data to infer a distance-dependent $v_{\rm los}$ field, but current applications are largely limited to the Solar neighborhood ($\sim$2~kpc) where reliable 3D dust maps are available.
Another complementary direction is simulation-based inference, which can jointly constrain a three-dimensional density field and a non-axisymmetric velocity field by forward-modeling PPV data, with hydrodynamical simulations in barred Galactic potentials providing physically motivated priors (Baba et al., in prep.).

Overall, BIKD provides a practical map-level reconstruction for the inner Milky Way that propagates line-of-sight distance ambiguities and quantifies model dependence via ensemble marginalization.
It is complementary to kinetic tomography and simulation-based inference approaches that incorporate independent distance anchors or fully forward-model PPV data.
By recovering a bar-aligned, quadrant-asymmetric molecular-gas morphology, BIKD helps bridge the gap between the well-mapped stellar bar and the more uncertain gas distribution, opening a path to coherent star--gas dynamical interpretations through direct comparison with modern stellar maps.

\section*{Funding}
This research was supported by the Japan Society for the Promotion of Science (JSPS) under Grant Numbers 21K03633, 21H00054, 22H01259, 24K07095, and 25H00394.

\section*{Data availability} 
 The simulation snapshots are available upon request.

\begin{ack}
We sincerely thank the anonymous referee for their thoughtful and constructive comments, which helped improve the clarity and context of this paper.
Calculations, numerical analyses, and visualization were carried out on Cray XD2000 (ATERUI-III,) and computers at the Center for Computational Astrophysics, National Astronomical Observatory of Japan (CfCA/NAOJ). 
\end{ack}

\end{document}